\documentclass[12pt]{article}
\usepackage{epsfig}
\usepackage{color}

\def\plb#1{Phys.~Lett.~{\bf B#1}}
\def\npb#1{Nucl.~Phys.~{\bf B#1}}
\def\prl#1{Phys.~Rev.~Lett.~{\bf #1}}
\def\prd#1{Phys.~Rev.~{\bf D#1}}

\def\abs#1{\left|{#1}\right|}

\def\e{\epsilon}

\def\L{\lambda}

\def\la{\langle}
\def\ra{\rangle}

\def\e3{$\epsilon_3$}

\def\bsgam{$b\rightarrow s\gamma$ }
\def\ch2{$\chi^2$}

\def\co#1{{\ifmmode{\cal O}_{#1}\else${\cal O}_{#1}$\fi}}

\def\dltmh{$\Delta m_H^2 \;\;$}
\def\dltmb{$\Delta m_b$}
\def\mupos{$\mu > 0 \;\;$}

\newdimen\unit
\def\point#1 #2 #3{\vbox to0pt{\kern-#2\unit
  \hbox{\kern#1\unit#3}\vss}
 \nointerlineskip}

\newcommand{\be}{\begin{equation}}
\newcommand{\ee}{\end{equation}}
\newcommand{\bea}{\begin{eqnarray}}
\newcommand{\eea}{\end{eqnarray}}

\newcommand{\gev}{\mbox{ GeV}}

\begin{document}
\thispagestyle{empty}
\noindent
\begin{flushright}
        OSHTPY-HEP-T-02-001\\
        January 2002
\end{flushright}

\vspace{1cm}
\begin{center}
  \begin{Large}
  \begin{bf}
Yukawa Unification in SO(10)
\end{bf}
 \end{Large}
\end{center}
  \vspace{1cm}
    \begin{center}
T. Bla\v{z}ek$^\dagger$, R. Derm\' \i \v sek$^*$ and S. Raby$^*$\\
      \vspace{0.3cm}
\begin{it}
$^\dagger$Department of Physics,
University of Southampton,
Southampton, UK \\
                   and Faculty of Mathematics and Physics,
           Comenius Univ., Bratislava, Slovakia\\
$^*$Department of Physics, The Ohio State University,
174 W. 18th Ave., Columbus, Ohio  43210
\end{it}
  \end{center}
  \vspace{1cm}
\centerline{\bf Abstract}
\begin{quotation}
\noindent
 In simple SO(10) SUSY GUTs the top, bottom
and tau Yukawa couplings unify at the GUT scale.   A naive renormalization
group analysis, neglecting weak scale threshold corrections, leads to
moderate agreement with the low energy data.  However it is known that intrinsically large threshold corrections
proportional to $\tan\beta \sim m_t(M_Z)/m_b(M_Z) \sim 50$ can nullify these
$t, \; b$, $\tau$ mass predictions.   In this paper we turn the argument around.   Instead of predicting fermion masses, we use the constraint of Yukawa unification and the observed values $M_t,\;  m_b(m_b),\; M_\tau$ to
constrain SUSY parameter space.  We find a narrow region survives for
$\mu > 0$ with $\mu,\; M_{1/2} << m_{16}$, $A_0 \approx - 1.9 \; m_{16}$ and $m_{16} > 1200$ \gev.   Demanding Yukawa unification thus
makes definite predictions for Higgs and sparticle masses.  In particular
we find a light higgs with mass $m_h^0 = 114 \pm 5 \pm 3$ GeV and a light stop with $(m_{\tilde t_1})_{MIN} \sim 450$ GeV and $m_{\tilde t_1} <<
m_{\tilde b_1}$.   In addition, we find a light chargino and a neutralino LSP.  It is also significant that in this region of parameter space the SUSY contribution to the muon anomalous magnetic moment $a_\mu^{SUSY} < 16 \times 10^{-10}$.
\end{quotation}
\vfill\eject

\section{Introduction}

Grand unification with $SU(5),\; SO(10)$ or even partial unification
with $SU(4)_C \times SU(2)_L \times SU(2)_R$ explains the peculiar
standard model charge assignments of quarks and leptons and also the
observed family structure~\cite{guts,eft,so10}.  Gauge coupling
unification at a scale $M_G \sim 3 \times 10^{16} \; {\rm GeV}$ in supersymmetric grand unified theories [SUSY GUTs] fits the low energy data well~\cite{susygut,susygut2,gutexp}.  Moreover SO(10) SUSY GUTs have many profound features~\cite{so10}:
\begin{itemize}
\item All fermions in one family sit in one irreducible {\bf 16} dimensional
representation.
\item  The two Higgs doublets, necessary in the minimal SUSY standard model [MSSM], sit in one irreducible {\bf 10} dimensional representation.
\item  Right-handed neutrinos, necessary for a see-saw mechanism for neutrino
masses, are naturally included in the {\bf 16} dimensional
representation.

\end{itemize}

In addition, in the simplest version of SO(10) the third generation Yukawa couplings  are given by a single term in the superpotential
$W =  \L \; {\bf 16 \; 10 \; 16}$ resulting in Yukawa unification  $\L_t = \L_b = \L_\tau = \L_{\nu_\tau} \equiv \; {\bf \L}$.  Hence, like gauge
coupling unification,  there is
a prediction but this time for  $M_t = 180 \pm 15 \; {\rm GeV}$
with large $\tan\beta \sim 50$ (see for example, Anderson et al.~\cite{nothreshcorr}); in good agreement with the data.
Note, GUT scale threshold corrections to this Yukawa unification boundary condition are naturally small ( $ \leq 1$\% ), unlike the corrections to gauge coupling unification which can easily be several percent (see the Appendix
for a discussion of perturbative GUT scale threshold corrections to gauge
and Yukawa couplings).

This beautiful prediction is however severely weakened by potentially
large weak scale threshold corrections
proportional to $\tan\beta$~\cite{threshcorr,bpr,pierceetal}. The complete
set of one loop corrections is given by
\be  m_b(M_Z) =
\L_b(M_Z) \; \frac{v}{\sqrt{2}} \; cos\beta \; (1 + \Delta m_b^{\tilde g} +  \Delta m_b^{\tilde \chi^+} + \Delta m_b^{\tilde \chi^0}
+ \Delta m_b^{\log} +  \Delta m_b^{EW}) . \ee
The first three terms are SUSY mass insertion corrections.
The dominant contributions
\be \Delta m_b^{\tilde g} \approx  \frac{2 \alpha_3}{3 \pi} \;
\frac{\mu m_{\tilde g}}{m_{\tilde b}^2} \; tan\beta \hspace{.5in}
{\rm and} \hspace{.5in}
 \Delta m_b^{\tilde \chi^+} \approx \frac{\L_t^2}{16 \pi^2} \;
\frac{\mu A_t}{m_{\tilde t}^2} \; tan\beta \label{eq:gluino} \ee
can be as large as 50\%.  Note in most regions of SUSY parameter space
these two terms have opposite sign.  $\Delta m_b^{\tilde \chi^0}$ is
on the other hand small, O(-1\%).   The log term results from finite wave function renormalization of the bottom quark; it is positive, independent of $\tan\beta$ and the total from all sources is of order 6\%.
Finally $\Delta m_b^{EW}$, due to Higgs, $W$ and $Z$ exchange, is negligibly small, O(.5\%).   There are similar corrections to $m_\tau$ and $m_t$.  The chargino corrections $m_\tau^{\tilde \chi^+}$ are also proportional to $\tan\beta$, but are significantly smaller than $\Delta m_b^{\tilde \chi^+}$ since
typically we have $m_{\nu_{\tilde \tau}} >> m_{\tilde t}$.
Finally the corrections to $m_t$ are not proportional to  $\tan\beta$.
The complete set of corrections can be found in the papers
by Rattazzi and Sarid~\cite{threshcorr} and by Pierce et al.~\cite{pierceetal}.

In most regions of SUSY parameter space $\Delta m_b^{\tilde g}$ is dominant
and in our conventions $\Delta m_b^{\tilde g} > 0$ for $\mu > 0$.
The sign of $\mu$ is constrained by experiment; in particular
$b \rightarrow s \gamma$ and $a_\mu^{NEW}$ both favor $\mu > 0$.
The same one loop graphs with a photon or gluon insertion and outgoing $b$
replaced with $s$ contributes to $b \rightarrow s \gamma$.
The chargino term typically dominates and has opposite sign to the SM and charged Higgs contributions, thus reducing the branching ratio for $\mu > 0$.  This is necessary to fit the data
since the SM contribution is somewhat too big.  $\mu < 0$ would on the other
hand constructively add to the branching ratio and is problematic.
In addition, the recent measurement of the anomalous magnetic moment of the muon
$a_\mu^{NEW} = (g - 2)/2 = 43 \; (16) \times 10^{-10}$ also favors \mupos~\cite{muon}.\footnote{Note, recent theoretical reevaluations of the standard model contribution are now closer to experiment with
$a_\mu^{NEW} = 25.6 \; (16) \times 10^{-10}$~\cite{g2new}.}

In a recent letter~\cite{bdr} we showed that Yukawa unification with
$\mu > 0$, including the complete one loop threshold corrections, is only consistent with the data in a narrow region of SUSY parameter space with
$\mu, \; M_{1/2} \sim 100 - 500 \; {\rm GeV}$;  $A_0 \sim -1.9 \; m_{16}$;  $m_{10} \sim 1.4 \; m_{16}$ and $m_{16} > 1200$ \gev.  The parameters
$m_{16},\; m_{10}$ denote the soft SUSY breaking mass terms for squark and
slepton, Higgs multiplets, respectively.  Note the requirement of Yukawa unification thus dramatically constrains the SUSY particle spectrum and Higgs masses.

In this paper we present a more detailed analysis of the SUSY particle
spectrum and the allowed parameter range.   In addition to fitting electroweak
data and the top, bottom and $\tau$ masses, we also include constraints
from $b \rightarrow s \; \gamma$ and $B_s \rightarrow \mu^+ \; \mu^-$.
The latter constraints increase the predicted stop mass
and the mass of the CP odd Higgs $A^0$, the heavy CP even Higgs $H^0$ and charged Higgs $H^\pm$ at the expense of a small increase in \ch2.
We find a light CP even Higgs boson with mass $m_h^0 = 114 \pm 5 \pm 3$ GeV.
In addition in the region where $m_{16} < 2000$ GeV, we find a light chargino with mass $m_{\tilde \chi^+} \sim 120 - 240$ GeV, a neutralino LSP with mass $m_{\tilde \chi^0} \sim 75 - 160$ GeV, a light stop with mass $m_{\tilde t_1} \sim 450 - 540$ GeV $<< m_{\tilde b_1}$.  The first and second generation 
squark and slepton masses are of order $m_{16}$.
It is also significant that in this region of parameter space we find
$a_\mu^{SUSY} < 16 \times 10^{-10}$.  Note also some recent discussions of Yukawa unification and $a_\mu^{SUSY}$~\cite{baerferrandis,nathetc}.

It is well known that electroweak symmetry breaking with large $\tan\beta$
and $m_{16} >> M_{1/2}$
requires Higgs up/down mass splitting~\cite{ewsb}.  We find however that
the fits to third generation fermion masses are sensitive to the mechanism
used to split the Higgs masses.  In this paper we consider D term
and ``Just So" Higgs splitting (defined in the text).
We study the sensitivity of our results to small GUT
scale threshold corrections to Yukawa couplings.   Significantly larger
threshold corrections are needed for D term splitting versus the Just So case.

The paper is organized as follows.
In section \ref{sect:analysis} we discuss the analysis.
We give the results for the case of Just So Higgs splitting
in section \ref{sect:resultsjustso} and D term Higgs splitting in
section \ref{sect:dterm}.  The constraints of $b \rightarrow s \; \gamma$
and $B_s \rightarrow \mu^+ \; \mu^-$ and additional experimental tests are considered in section \ref{sect:tests}.
For the impatient reader we present detailed results from some typical points in SUSY parameter space in Table 1 (without) and Table 2 (with) the constraints from $b \rightarrow s \; \gamma$ and $B_s \rightarrow \mu^+ \; \mu^-$ included.
Note, we have included the prediction for $a_\mu^{SUSY}$ in Tables 1 and 2; however it has not been included in the \ch2 function when fitting.
Finally some theoretical questions are addressed in section \ref{sect:discussion}.

\section{Analysis}
\label{sect:analysis}
We use a top - down approach with a global \ch2 analysis~\cite{chi2}.
The input parameters are defined by boundary conditions at the GUT scale.
The 11 input parameters at $M_G$ are given by --- three gauge parameters
$M_G, \; \alpha_G(M_G),$ $\epsilon_3$; the Yukawa coupling
$\L$, and 7 soft SUSY breaking parameters
$\mu,\; M_{1/2},\; A_0,\;  \tan\beta$;  $m_{16}^2, \; m_{10}^2, $ \dltmh $\; (D_X)$ for Just So (D term) case.\footnote{$\epsilon_3$, defined in the Appendix, and \dltmh, $D_X$ parametrize GUT scale
threshold corrections to gauge coupling unification  and Higgs up/down mass
splitting, respectively.}
These are fit in a global \ch2 analysis defined in terms of physical low
energy observables. We use two (one)loop renormalization group [RG] running
for dimensionless (dimensionful) parameters from $M_G$ to
$M_Z$.  We require electroweak symmetry breaking using an improved Higgs potential, including $m_t^4$ and $m_b^4$ corrections in an effective 2
Higgs doublet model below $M_{stop}$~\cite{carenaetal}.  The \ch2 function includes 9 observables; 6 precision electroweak data $\alpha_{EM},$
$G_\mu,$  $\alpha_s(M_Z),$ $M_Z, \; M_W,$ $\rho_{NEW}$ and the 3 fermion
masses $M_{top},\;  m_b(m_b), \; M_\tau$.\footnote{Capital $M$ is used
for pole masses and lower case $m$ for $\overline{MS}$ running masses.}  We fit the central values:~\cite{pdg2000}
$M_Z = 91.188$ GeV, $M_W = 80.419$ GeV, $G_{\mu}\times 10^5  = 1.1664$ GeV$^{-2}$, $\alpha_{EM}^{-1} = 137.04,$ $M_{\tau} = 1.7770$ GeV with
0.1\% numerical uncertainties; and the following with the experimental
uncertainty in parentheses:
$\alpha_s(M_Z) = 0.1180\; (0.0020),$ $\rho_{new}\times 10^3 =
-0.200\; (1.1)$~\cite{rhonew}, $M_t  = 174.3\; (5.1)$ GeV,
$m_b(m_b)  = 4.20\; (0.20)$ GeV.\footnote{The error for $m_b(m_b)$~\cite{pdg2000}
appears to be quite conservative in view of recent claims to much
smaller error bars~\cite{bmass}.}
We include the complete one loop threshold corrections at $M_Z$ to all
observables.  In addition we use one loop QED and three loop QCD RG running below $M_Z$.  Finally, with regards to the calculated
Higgs and sparticle masses,
the neutral Higgs masses $h,\; H, \; A^0$ are pole masses calculated with
the leading top, bottom, stop, sbottom loop contributions; while all other
sparticle masses are running masses.

 We minimize \ch2 using
the CERN subroutine minuit.  In order to present our results we typically keep
three parameters (such as $\mu,\; M_{1/2},\; m_{16}$) fixed and minimize
\ch2 with respect to the remaining eight parameters.   We then plot our
results as contours in the two parameter space.

\subsection{EWSB and Higgs mass splitting}
The first significant constraint derives from electroweak symmetry
breaking [EWSB] in the large $\tan\beta$ regime.   It has been shown
that this typically requires
$m_{H_u}^2 <  m_{H_d}^2$.  In fact more general solutions for EWSB exist
with Higgs up/down splitting and with less fine-tuning (see ~\cite{ewsb} and Rattazzi and Sarid~\cite{threshcorr}). The range of soft SUSY parameters
required is consistent with solution (B) of Olechowski and Pokorski~\cite{ewsb}.

 In our analysis we consider two particular Higgs splitting
schemes, we refer to as Just So and D term splitting.
In the first case the third generation squark and slepton soft
masses are given by the universal mass parameter $m_{16}$ , and
only Higgs masses are split:  {\bf Just So Higgs splitting}
\be
m_{(H_u, \; H_d)}^2 = m_{10}^2 \;( 1 \mp \Delta m_H^2) \ee
In this case we find \dltmh $\sim 13$ \%.

In the second case we assume
D term splitting, i.e. that the D term for $U(1)_X$ is non-zero,
where $U(1)_X$ is obtained in the decomposition of $SO(10)
\rightarrow SU(5) \times U(1)_X$.  In this second case, we have:
{\bf D term splitting}
\bea m_{(H_u,\; H_d)}^2 =  m_{10}^2 \mp 2 D_X ,& &  \label{eq:dterm} \\
 m_{(Q,\; \bar u,\; \bar e)}^2 =  m_{16}^2 + D_X, & & \nonumber \\
m_{(\bar d,\; L)}^2 =  m_{16}^2 - 3 D_X . \nonumber \eea
Here we find $\Delta m_H^2 \equiv 2 \ D_X/m_{10}^2 \sim 5$\%.

Just So Higgs splitting does not at first sight appear to be as well 
motivated as D term splitting.  In the Appendix we present two example mechanisms for Just So Higgs splitting.  Here we present the most compelling argument.  In $SO(10)$, neutrinos necessarily have a Yukawa term 
coupling active neutrinos to the ``sterile" neutrinos present in the {\bf 16}.
In fact for $\nu_\tau$ we have $\L_{\nu_\tau} \; \bar \nu_\tau \; L \; H_u$ 
with $\L_{\nu_\tau} = \L_t = \L_b = \L_\tau \equiv \; {\bf \L}$.   
In order to obtain a tau neutrino mass with $m_{\nu_\tau} \sim 0.06$ eV (consistent with atmospheric neutrino oscillations), the ``sterile" $\bar \nu_\tau$ must obtain  a Majorana mass $M_{\bar \nu_\tau} \geq 10^{13}$ GeV.   Moreover, since neutrinos couple to $H_u$ (and not to $H_d$) with a fairly large Yukawa coupling (of order 0.7), they naturally distinguish the two Higgs multiplets.  With $\L = 0.7$ and $M_{\bar \nu_\tau} = 10^{13}$ GeV, we obtain a 
significant GUT threshold correction with $\Delta m_H^2 = .10$, remarkably close
to the value needed to fit the data.   At the same time, we obtain a small
threshold correction to Yukawa unification $< 3$\%.  (For more details, see
the Appendix.)

\section{Results: Just So Higgs splitting}
\label{sect:resultsjustso}

Since the log corrections $\Delta m_b^{\log} \sim \; $O(6\%)  are positive,
they must be cancelled in order to obtain \dltmb $\leq - 2$ \% to
fit $m_b$.   For \mupos the gluino contribution is positive.
The chargino contribution is typically opposite in sign to the gluino,
since $A_t$ runs to an infrared fixed point, $A_t \propto - M_{1/2}$ (see for example, Carena et al.~\cite{threshcorr}).  Hence in order to cancel the positive
contribution of both the log and gluino contributions, a large negative
chargino contribution is needed.   This can be accomplished for
$- A_t > m_{\tilde g}$ and  $m_{\tilde t} << m_{\tilde b}$.  The first
condition can be satisfied for $A_0$ large and negative, which helps pull
$A_t$ away from its infrared fixed point.   The second condition is also
aided by large $A_t$.  However in order to obtain a large enough splitting between $m_{\tilde t}$ and $m_{\tilde b}$, large values of $m_{16}$ are needed.    Note, that for universal scalar masses, the lightest stop is typically
lighter than the sbottom.  On the other hand, D term splitting with
$D_X > 0$ gives $m_{\tilde b} \leq m_{\tilde t}$.  Recall $D_X > 0$ is needed
for electroweak symmetry breaking.  As a result in the case
of Just So boundary conditions excellent fits are obtained for the top,
bottom and tau masses; while for D term splitting the best fits give
$m_b(m_b) \geq 4.59 \; {\rm GeV}$.   In this section we give the results
for the case of Just So Higgs splitting.  The results for D term splitting are discussed in section \ref{sect:dterm}.

\begin{figure}
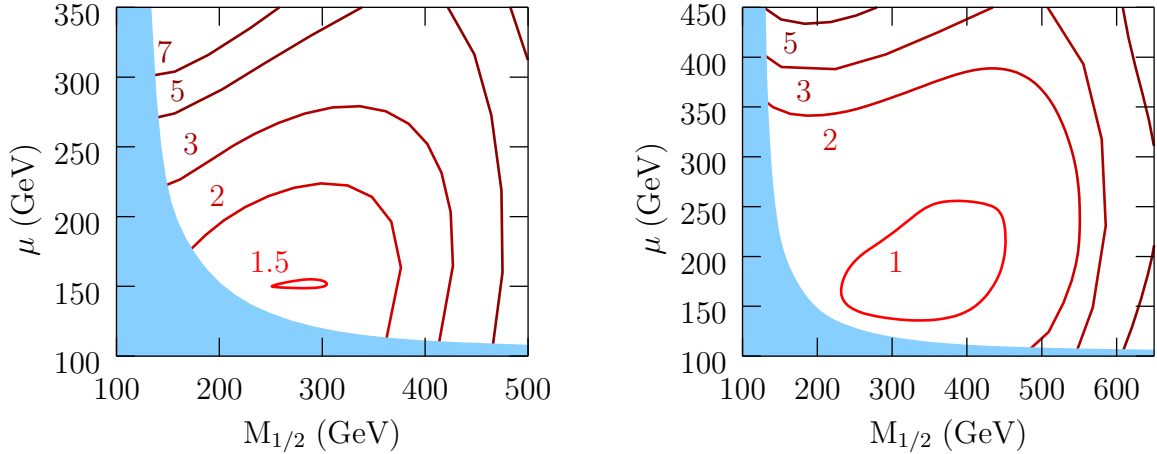

\begin{center}
\begin{tabular}{cc}
\input{m_1500_chi2.pstex_t}
\label{figure:chi2}   &
\hspace{.5cm}  \input{m_2000_chi2.pstex_t}
\end{tabular}
\end{center}
\caption{$\chi^2$ contours for $m_{16} = 1500$ GeV (Left) and $m_{16} = 2000$ GeV (Right).}
\end{figure}

\begin{figure}
\begin{center}
\centerline{\input{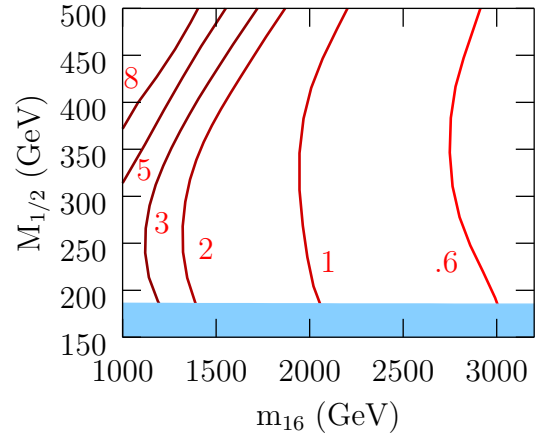}}
\end{center}
\caption{$\chi^2$ contours for $\mu = 150$ GeV.}
\end{figure}

In Fig. 1, we show \ch2 contours as a function of $\mu, \; M_{1/2}$
for $m_{16} =   (1500) \; 2000 \; {\rm GeV}$.  The shaded region {\em in
all the figures} is excluded by the experimental bound on the chargino mass,
 $m_{\tilde \chi^+} > 103 \; {\rm GeV}$.
The \ch2 $< 2$ contour for $m_{16} = 1500 \; {\rm GeV}$  is bounded
by  ${\rm shaded} < \mu < 220 \; {\rm GeV}$,  ${\rm shaded} < M_{1/2} < 380 \; {\rm GeV}$.   For $m_{16} = 2000 \; {\rm GeV}$ the region with \ch2 $< 1 (2)$
is contained within the closed curve bounded by $150 \; ({\rm shaded}) < \mu <$
$250 \; (380) \; {\rm GeV}$,  $220 \; ({\rm shaded}) < M_{1/2} < 450 \; (550)\; {\rm GeV}$.
In Fig. 2, we show \ch2 contours as a function
of   $M_{1/2}, \; m_{16}$ for $\mu = 150 \; {\rm GeV}$; \ch2 $< 1$ for
$m_{16} \geq 2000 \; {\rm GeV}$.   We see \ch2 continues to decrease as $m_{16}$ increases.

In Table 1 (Fits 1,2) we present the input parameters and output for
two representative points with universal squark and slepton masses
$m_{16} = 1500 \; (2000) \; {\rm GeV}$ with  $\mu = 150 \; (200)\; {\rm GeV}, \;
M_{1/2} = 250 \; (350) \; {\rm GeV}$.  We find reasonable fits (\ch2 $\leq 3$) only for $m_{16} \geq 1200 \; {\rm GeV}$.  For $m_{16} < 1200 \; {\rm GeV}$,
\ch2 increases rapidly.

The bottom quark mass $m_b(m_b)$ is given in Fig. 3 (Left) for $m_{16} = 2000$ GeV.  In Fig. 3 (Right) we show that the fits improve with good fits extending to larger values of $M_{1/2}$ as $m_{16}$ increases ($\mu
= 150$ GeV is fixed).  It should be clear that $m_b$ is the dominant pull on \ch2 as seen by comparing to the \ch2 contours of Fig. 1 (Right).

\begin{figure}
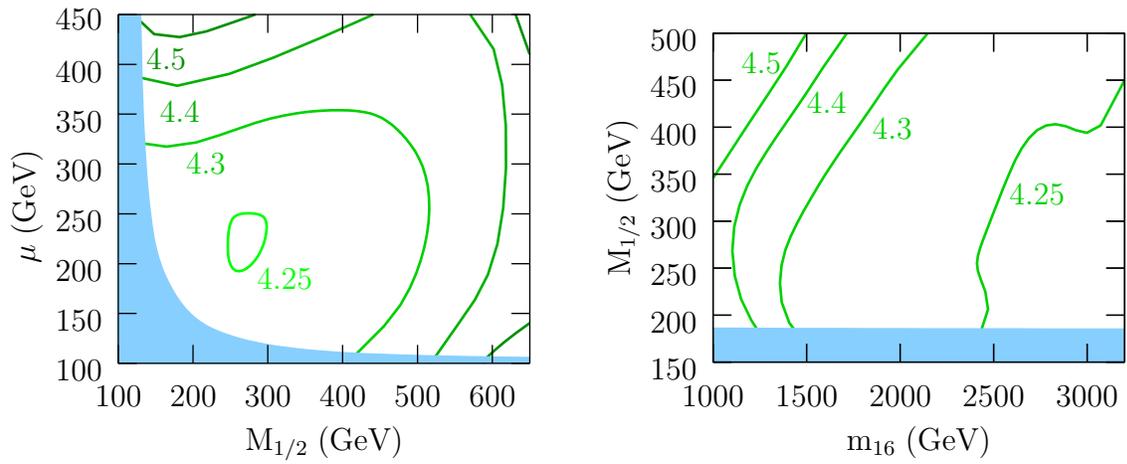

\begin{center}
\begin{tabular}{cc}
\input{m_2000_m_b.pstex_t}
\hspace{.5cm}  \input{m_M_mb.pstex_t}
\end{tabular}
\end{center}
\caption{Contours of constant $m_b(m_b)$ [GeV] for $m_{16} = 2000$ GeV (Left) and for $\mu = 150$ GeV
(Right).}   
\end{figure}

\begin{figure}
\begin{center}
\input{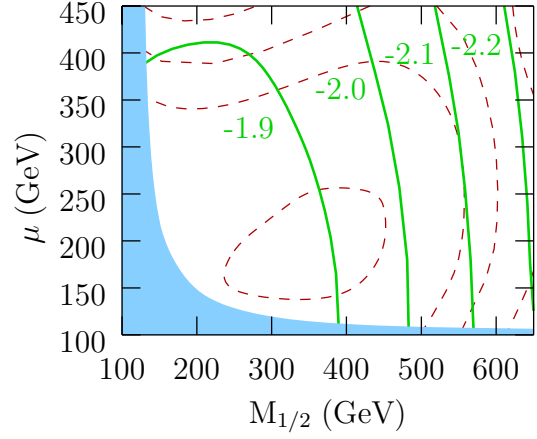}
\end{center}
\caption{$A_0/m_{16}$ contours for $m_{16} = 2000$ GeV, with \ch2 contours
overlayed.} 
\end{figure}

\begin{figure}
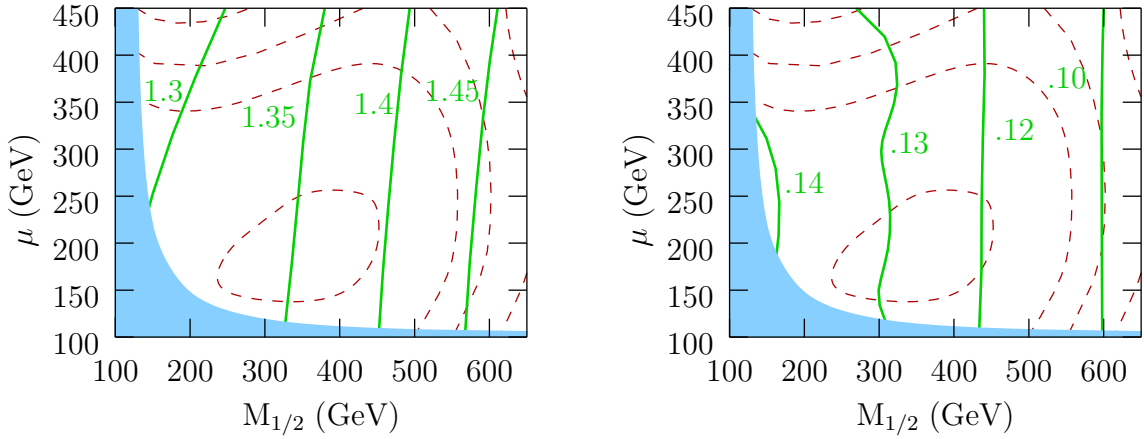

\begin{center}
\begin{tabular}{cc}
\input{m_2000_m10_over_m16_chi2.pstex_t} &
\hspace{.5cm}
\input{m_2000_delta_mh2_chi2.pstex_t}
\end{tabular}
\end{center}
\caption{$m_{10}/m_{16}$ contours (Left) and $\Delta m_H^2$ contours (Right)
for $m_{16} = 2000$ GeV, with \ch2 contours overlayed.}
\end{figure}

In Fig. 4 we plot $A_0/m_{16}$ at $M_G$ as a function of
$\mu, \; M_{1/2}$ for fixed  $m_{16} = 2000 \; {\rm GeV}$.
Good fits are obtained for $A_0 \approx - 1.9 \; m_{16}$ for all
$m_{16} > 1200$ GeV.
Note, even though $A_0$ is very large, the value of $A_t$ at $M_Z$ is significantly smaller since it is driven to an infra-red quasi
fixed point.  [We  come back to this point when we discuss vacuum
stability issues later.]  Finally  reasonable fits require $m_{10} \sim
1.35 \; m_{16}$ (Fig. 5 Left) and $\Delta m_H^2 \sim 0.13$ (Fig. 5 Right).

The significant positive log correction to $m_b$ is the main
reason why Yukawa unification is only possible in a narrow region of SUSY parameter space.  In order to compensate this, the chargino mass insertion contribution must be significantly larger than the gluino contribution.
In Fig. 6 we give the gluino (Left), chargino (Right) mass insertion corrections to $m_b$ and the total 
weak scale threshold correction $\delta m_b$ (lower Center) for fixed $m_{16} = 2000 \; {\rm GeV}$ as a function
of $\mu, \; M_{1/2}$.   Note in the region of \ch2 $< 1$ the gluino (chargino)
mass insertion corrections are large and of order 13 to 26 \% (- 23 to -34 \%),
while the log correction is 5.6 to 6.6 \%.   These are the dominant
corrections.  The total SUSY correction to $m_b$ is -3 to - 4\%.
In the same region, the total SUSY correction to $M_t$ is 7 - 8 and to
$M_\tau$ is - 2 to - 4\%.

\begin{figure}
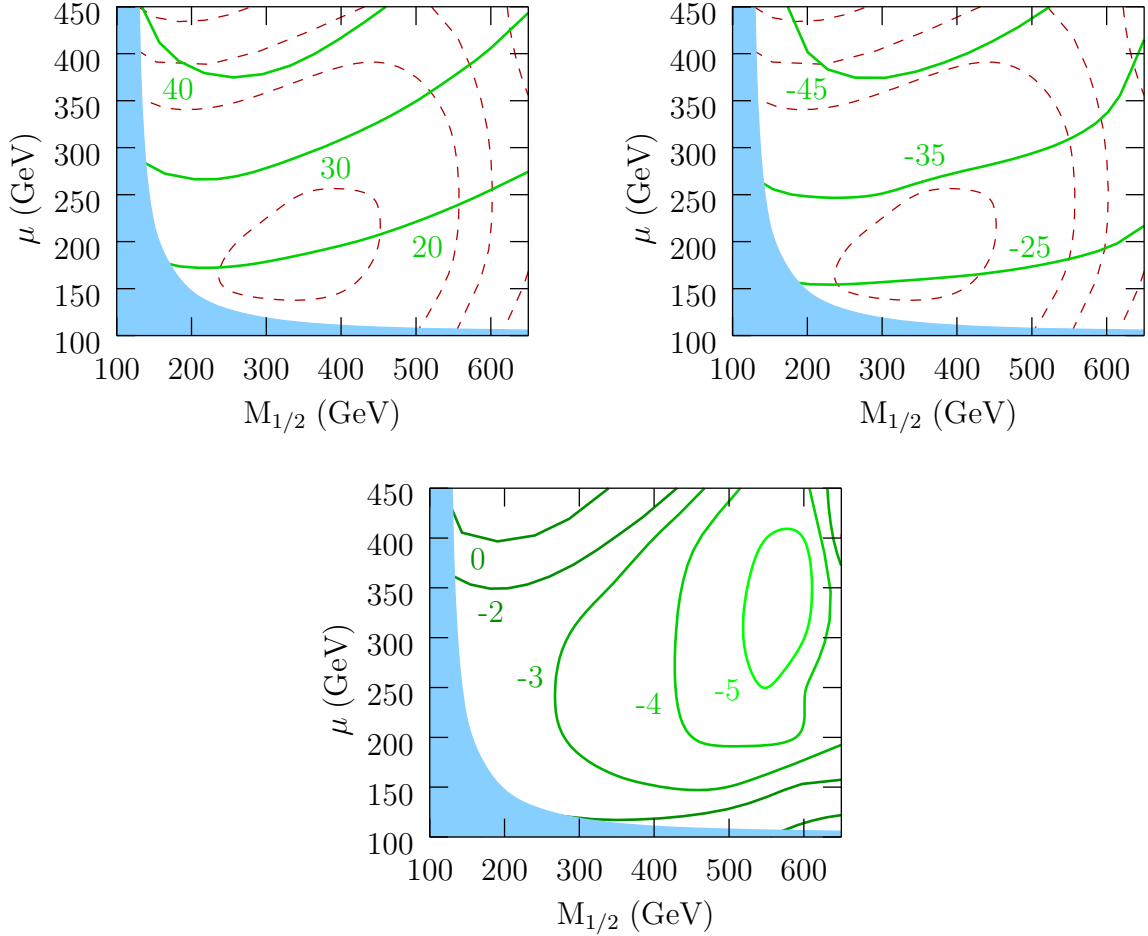

\begin{center}
\begin{tabular}{cc}
\input{m_2000_gl_mass_corr_chi2.pstex_t} &
\hspace{.5cm}  \input{m_2000_ch_mass_corr_chi2.pstex_t}
\end{tabular}
\end{center}
\begin{center}
\input{m_2000_m_b_corr.pstex_t}
\end{center}
\caption{Contours of constant mass insertion corrections to $\delta m_b$ [\%] from gluino loops (Left), chargino loops 
(Right) (with \ch2 contours overlayed) and the total one loop correction $\delta m_b$ [\%] (lower Center) 
for $m_{16} = 2000$ GeV.} 
\end{figure}

In summary, we have shown that good fits to $b, t$ and $\tau$ masses are only obtained in a narrow region of SUSY parameter space  $A_0 \approx - 1.9 \; m_{16}$, $m_{10} \sim  1.35 \; m_{16}$  with $m_{16} > 1200$ GeV.  A Just So
Higgs splitting $\Delta m_H^2 \sim 0.13$ is also required.\footnote{Note
in our analysis we have fit $\alpha_s(M_Z) = 0.118 \pm 0.002$.
However if the central value for $\alpha_s(M_Z)$ were to
decrease to 0.116 we would obtain good fits for $m_b(m_b)$ in a larger
region of SUSY parameter space and still have $|\epsilon_3| < 5$\%.  This is because both RG running and the
gluino correction to the bottom mass are positive and proportional to
$\alpha_s(M_Z)$.}

This has interesting consequences for the Higgs and supersymmetric particle spectrum.
In Figs. 7 - 9 we give the $A^0, \; h^0,\; H^0,\; H^\pm$ masses.
In Fig. 7,  constant $m_{A^0}$ contours are given
 as a function of $\mu, \; M_{1/2}$ for fixed  $m_{16} = 2000 \;
{\rm GeV}$.  We find, for \ch2 $< 1$, $m_{A^0} \sim 106 - 112$ GeV.
We show constant $m_{h^0}$ contours for $\mu, \; M_{1/2}$ for fixed
$m_{16} = 1500$ GeV in Fig. 8 (Left), for $m_{16} = 2000 \; {\rm GeV}$ (Right)
and as a function of $M_{1/2}, \; m_{16}$ for $\mu = 150$ GeV in Fig. 8 (lower Center).   We find, for \ch2 $< 1$, $m_h^0 \sim 112 - 117$ GeV.
Finally, in Fig. 9, constant $m_{H^0}$ and $m_{H^\pm}$ contours are given as a function of $\mu, \; M_{1/2}$ for fixed  $m_{16} = 2000 \;
{\rm GeV}$.   We find, for \ch2 $< 1$,
$m_H^0 \sim  118 - 121  $ GeV and $m_H^\pm \sim  145 - 149 $ GeV.

\begin{figure}
\begin{center}
\input{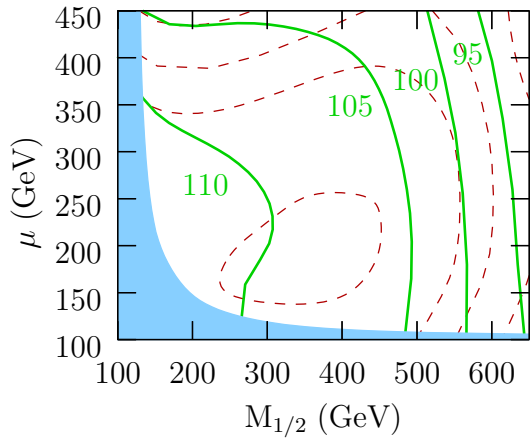}
\end{center}
\caption{Contours of constant $A^0$ mass [GeV] with fixed $m_{16} = 2000$ GeV.}
\end{figure}

\begin{figure}
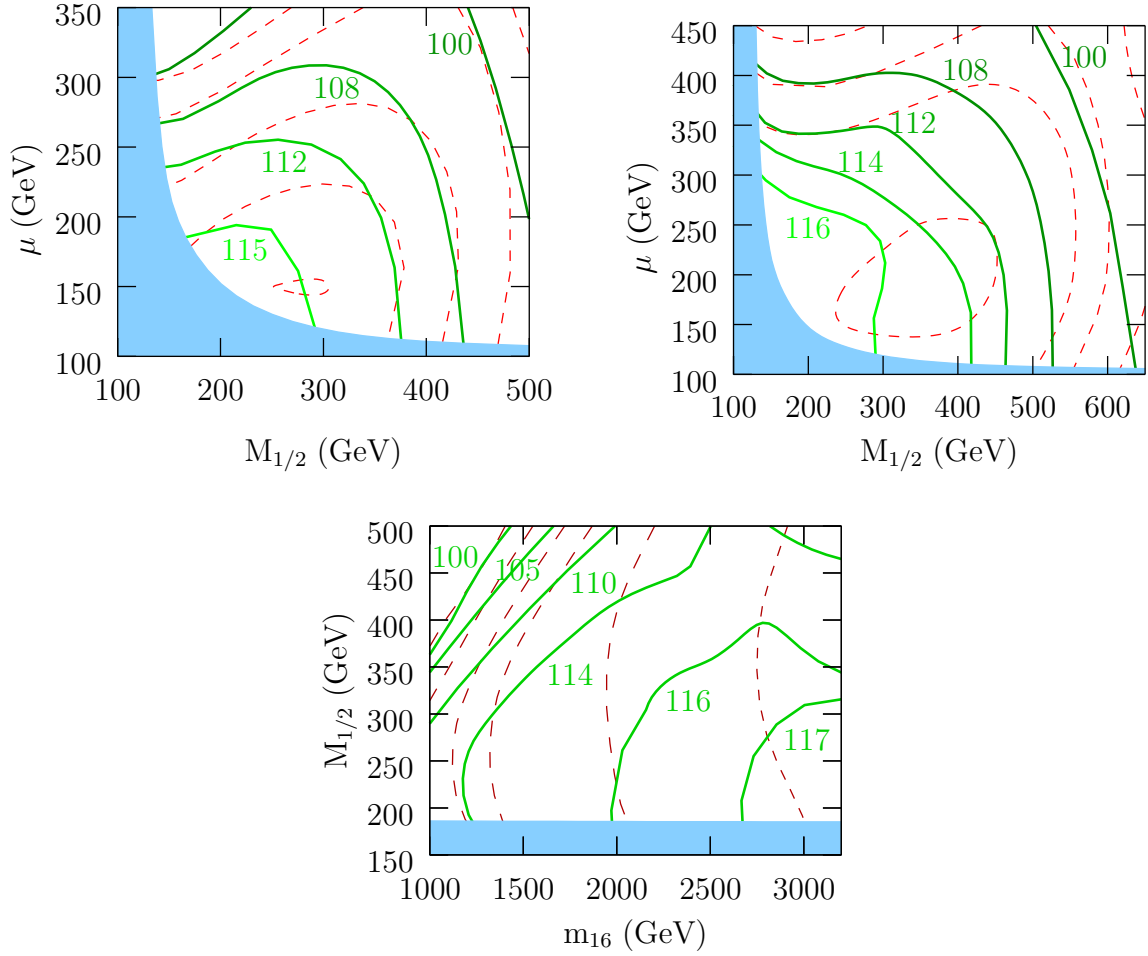

\begin{center}
\begin{tabular}{cc}
\input{m_1500_h0.pstex_t} &
\hspace{.5cm}  \input{m_2000_h0.pstex_t}
\end{tabular}
\end{center}
\begin{center}
\input{m_M_h0_chi2.pstex_t}
\end{center}
\caption{Contours of constant $h^0$ mass [GeV] with fixed $m_{16} = 1500$
GeV (Left); $m_{16} = 2000$ GeV (Right)
and with fixed $\mu = 150$ GeV as a function of $M_{1/2},\;  m_{16}$ (lower
Center).}
\end{figure}

In Figs. 10 - 13 we show constant mass contours for $\tilde t_1,\; \tilde t_2,\; \tilde b_1,$ and $\tilde \tau_1$ for fixed $m_{16} = 2000$ GeV.
We find for \ch2 $< 1$,  $\tilde t_1 \sim 175 - 250$ GeV,  $\tilde t_2 \sim 630 - 800 $ GeV, $\tilde b_1 \sim 500 - 650$ GeV,  $\tilde \tau_1 \sim  250 - 500 $ GeV.  Note, the upper bounds on squark and slepton masses increase
as $m_{16}$ increases.  Moreover, the first two generation squark and
slepton masses are of order $m_{16}$.

Gaugino masses are smooth functions of $\mu, \; M_{1/2}$.   The gluino mass is linear in $M_{1/2}$ and 
satisfies the empirical relation $m_{\tilde g} = 2.5 \; M_{1/2} + 25 \; {\rm GeV}.$   In Fig. 14 we 
show constant mass contours for $\tilde \chi^\pm_1$ and for $\tilde \chi^0_1$, the LSP. 
We find, for \ch2 $< 1$,
 $\tilde \chi^\pm_1 \sim 120 - 240  $ GeV and
 $\tilde \chi^0_1 \sim 75 - 160 $ GeV.
Finally the GUT scale parameters $M_G, \; \alpha_G(M_G),$
$\epsilon_3$, and $\L \sim .65 - .7$ and weak scale parameter
$\tan\beta \sim 50 - 52$ are weakly dependent on SUSY parameters.

\begin{figure}
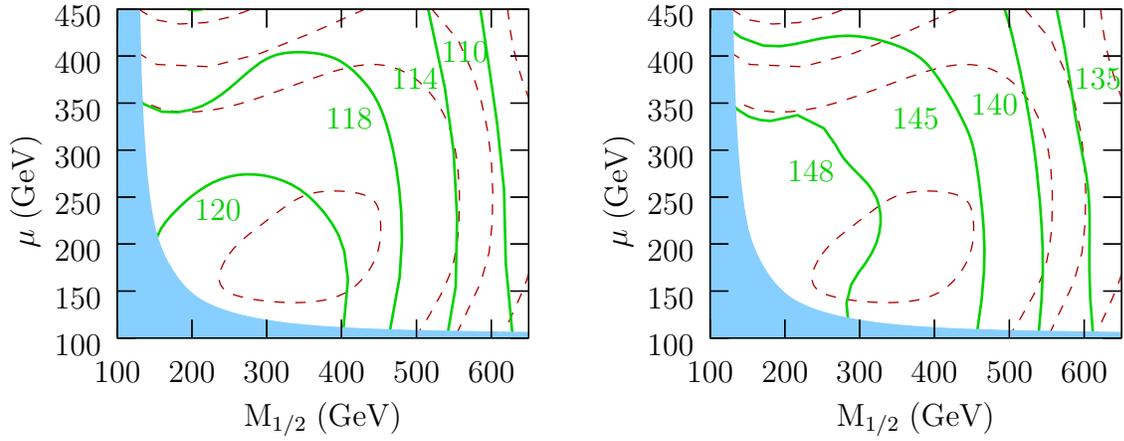

\begin{center}
\begin{tabular}{cc}
\input{m_2000_H0_chi2.pstex_t}
\hspace{.5cm}  \input{m_2000_Hpm_chi2.pstex_t}
\end{tabular}
\end{center}
\caption{Contours of constant $H^0$ mass [GeV] (Left) and $H^\pm$ mass [GeV] (Right)
with fixed $m_{16} = 2000$ GeV.}
\end{figure}

\begin{figure}
\begin{center}
\begin{tabular}{cc}
\input{m_2000_stop1_chi2.pstex_t} &
\hspace{.5cm}  \input{m_M_stop1_chi2.pstex_t}
\end{tabular}
\end{center}
\caption{Contours of constant $\tilde t_1$ mass [GeV] for fixed $m_{16} = 2000$ GeV (Left) and  fixed $\mu = 150$ GeV 
(Right) with constant \ch2 contours overlayed.}
\end{figure}

\begin{figure}
\begin{center}
\begin{tabular}{cc}
\input{m_2000_stop2_chi2.pstex_t} &
\hspace{.5cm}  \input{m_M_stop2_chi2.pstex_t}
\end{tabular}
\end{center}
\caption{Contours of constant $\tilde t_2$ mass [GeV]  for fixed $m_{16} = 2000$ GeV (Left) and  fixed $\mu = 150$ GeV 
(Right) with constant \ch2 contours overlayed.}
\end{figure}

\begin{figure}
\begin{center}
\begin{tabular}{cc}
\input{m_2000_sbottom1_chi2.pstex_t}  &
\hspace{.5cm}  \input{m_M_sbottom1_chi2.pstex_t}
\end{tabular}
\end{center}
\caption{Contours of constant $\tilde b_1$ mass [GeV] for fixed $m_{16} = 2000$ GeV (Left) and  fixed $\mu = 150$ GeV 
(Right) with constant \ch2 contours overlayed.}
\end{figure}

\begin{figure}
\begin{center}
\begin{tabular}{cc}
\input{m_2000_stau1_chi2.pstex_t}  &
\hspace{.5cm}  \input{m_M_stau1_chi2.pstex_t}
\end{tabular}
\end{center}
\caption{Contours of constant $\tilde \tau_1$ mass [GeV] for fixed $m_{16} = 2000$ GeV (Left) and  fixed $\mu = 150$ 
GeV (Right) with constant \ch2 contours overlayed.}
\end{figure}

\begin{figure}
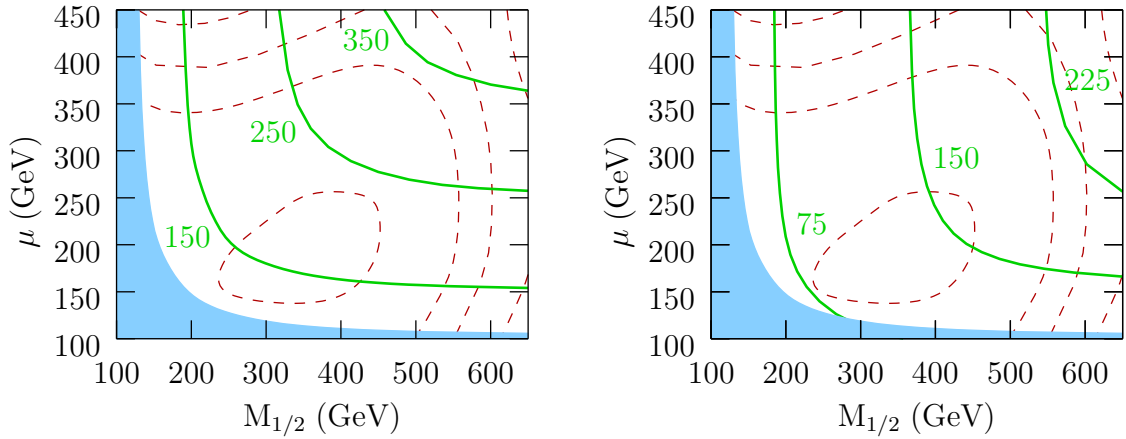

\begin{center}
\begin{tabular}{cc}
\input{m_2000_chargino_chi2.pstex_t}
\hspace{.5cm}  \input{m_2000_neutralino1_chi2.pstex_t}
\end{tabular}
\end{center}
\caption{Contours of constant $\tilde \chi^\pm_1$ mass [GeV] (Left)
and constant $\tilde \chi^0_1$ mass [GeV] (Right)
for fixed $m_{16} = 2000$ GeV with constant \ch2 contours overlayed.}
\end{figure}

In the following we evaluate the sensitivity of our results to plausible
threshold corrections to Yukawa unification at $M_G$.  We consider
two loop RG running of dimensionful parameters.  We also artificially
fix the CP odd Higgs mass $m_{A^0}$ by applying appropriate penalties to
the \ch2 function.  We then discuss the dependence of the Higgs spectrum
and \ch2 as a function of $m_{A^0}$.  This will become important when
considering the decay $B_s \rightarrow \mu^+ \; \mu^-$ in section \ref{sect:tests}.
In addition in \ref{sect:tests}
we discuss constraints
from the process $b \rightarrow s \; \gamma$.   Both of these latter processes
require a description of Yukawa matrices for the heaviest two families.

\subsection{Sensitivity to GUT scale threshold corrections}

\begin{figure}
\begin{center}
\input{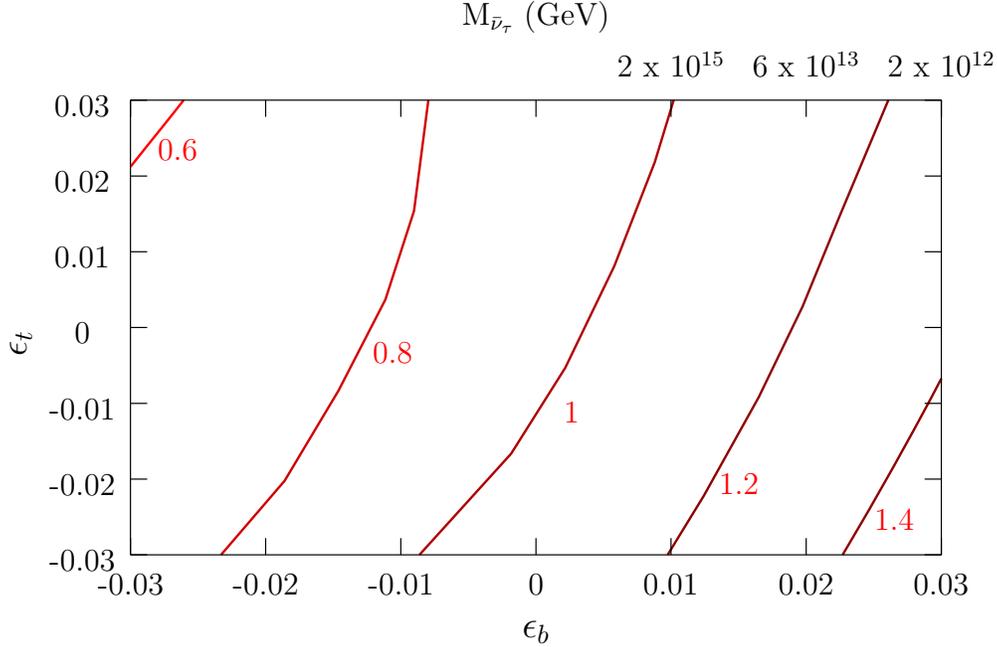}
\end{center}
\caption{\ch2 contours as a function of $\epsilon_b, \epsilon_t$ for $M_{1/2} = 300 \;{\rm GeV},\; \mu = 150 \; {\rm
GeV}, m_{16} = 2 \;{\rm TeV}$.   Just So scalar mass case.  On the upper axis we give the equivalent value of
$M_{\nu_\tau} << M_G$ (the Majorana mass of the tau neutrino) which contributes to $|\epsilon_b|$ with $|\epsilon_t| =
0$.}
\end{figure}

In this section we check the sensitivity of our results to GUT scale
threshold corrections to Yukawa unification.   We define $\epsilon_b, \epsilon_t$ by \be \lambda_i = \lambda \; ( 1 + \epsilon_i ) \;\;\; {\rm with}
\;\;\; i = b,t  \;\;\; {\rm and} \;\;\; \lambda_\tau \equiv \lambda .\ee
In Fig. 15 we give \ch2 contours as a function of $\epsilon_b, \epsilon_t$
for $M_{1/2} = 300 \;{\rm GeV},\; \mu = 150 \; {\rm GeV}, m_{16} = 2000 \;{\rm
GeV}$.   We consider values of $|\epsilon_b|, |\epsilon_t| < 3$\%.
It is clear that \ch2 is not very sensitive to small Yukawa threshold
corrections.

The best motivated correction comes from integrating out a heavy
tau neutrino with mass $M_{\nu_\tau}$.
Neutrino oscillations consistent with atmospheric neutrino data
suggest $M_{\bar \nu_\tau} \approx 10^{13} \; {\rm GeV}$ corresponding
to a correction $\epsilon_b = 2.6$\% with $|\epsilon_t| = 0$.
On the upper axis, in Fig. 15, we give the equivalent value of
$M_{\nu_\tau} << M_G$ (the Majorana mass of the tau neutrino) which
contributes to $|\epsilon_b|$ with $|\epsilon_t| = 0$ (see the Appendix).
In section \ref{sect:dterm} we consider D term Higgs splitting where
threshold corrections are absolutely essential for reasonable fits.

\subsection{Two loop vs. one loop RGEs}
In the region of parameter space we consider, with $m_{16} >$ TeV,
two loop RG running of soft SUSY breaking parameters may have significant
consequences for sparticle masses as well as for electroweak symmetry breaking.
We have checked however that a two loop RGE analysis for soft SUSY masses
does not significantly affect our results.  By this we mean that the same
low energy results are obtained with small changes in the GUT scale
parameters.  In Table 1 (Fit 3) we present the comparison for $m_{16} =
2000$\gev, $\mu = 150 \; {\rm GeV}, \; M_{1/2} = 300 \;
{\rm GeV}$ fixed with all other input parameters varied to minimize
\ch2 using one and two loop RGEs.  It is clear that the results are
not significantly different from Fit 2.  A small change in $A_0$ is
sufficient to guarantee positive squark masses squared.
Of course, when one uses two loop RGEs for soft scalar masses
consistency requires including one loop threshold corrections to these masses
at the weak scale.   We have not included the latter contributions in our
calculations; thus we stick with the one loop RGE analysis from $M_G$ to
$M_Z$ for dimensionful parameters.

\subsection{$\chi^2$ dependence on $m_{A^0}$ mass}

In the course of our analysis it became clear that there were
two minima for \ch2; with a low and high mass solution for the CP odd
Higgs mass (see also Ref. \cite{br}).  In order to make this behavior explicit we needed a way
to choose particular values of $m_{A^0}$ within the \ch2 analysis.
We accomplished this by adding a penalty to the \ch2 function for any value
of $m_{A^0}$ outside a narrow range.  Note, we then found minima of \ch2
for which this penalty vanished.

\begin{figure}
\begin{center}
\input{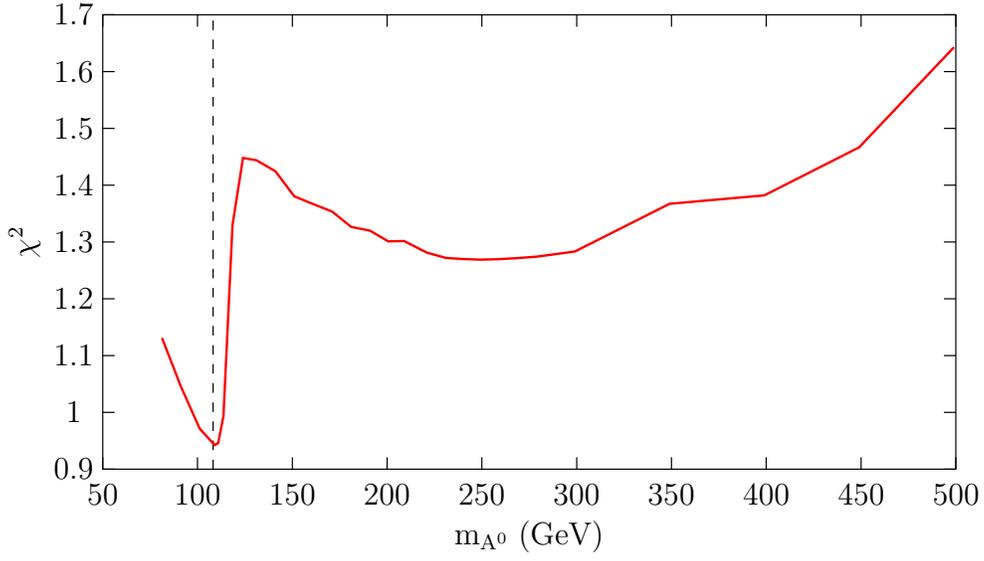}
\end{center}
\caption{\ch2 as a function of the CP odd Higgs mass $m_A^0$.}
\end{figure}

In Fig. 16 we plot \ch2 as a function of $m_{A^0}$ for fixed
$m_{16} = 2000$ GeV, $\mu = 150 \; {\rm GeV}, \; M_{1/2} = 300 \;
{\rm GeV}$.   The global minimum is at $m_{A^0} = 110$ GeV
with the local minimum at $m_{A^0} \sim 250$ GeV with approximately
 35\% larger \ch2.
The increased pull to \ch2 is mainly due to the $\rho$ parameter.
In Fig. 17 we plot the light Higgs mass vs. $m_{A^0}$.  Note, at the
minimum of \ch2, $m_{h^0}$ is a steeply rising function of $m_{A^0}$.
However it quickly reaches a plateau with $m_{h^0} \sim 119$\gev.
In Fig. 18 we see that $m_{H^0}$ and $m_{H^\pm}$ increase linearly with $m_{A^0}$.  We conclude therefore that the non-SM Higgs masses
are not constrained by Yukawa unification.  Nevertheless,  
all the other predictions (the region of SUSY parameter
space, the light higgs mass and the sparticle spectrum) remain 
unchanged, i.e. independent of the non-SM Higgs masses.

\begin{figure}
\begin{center}
\input{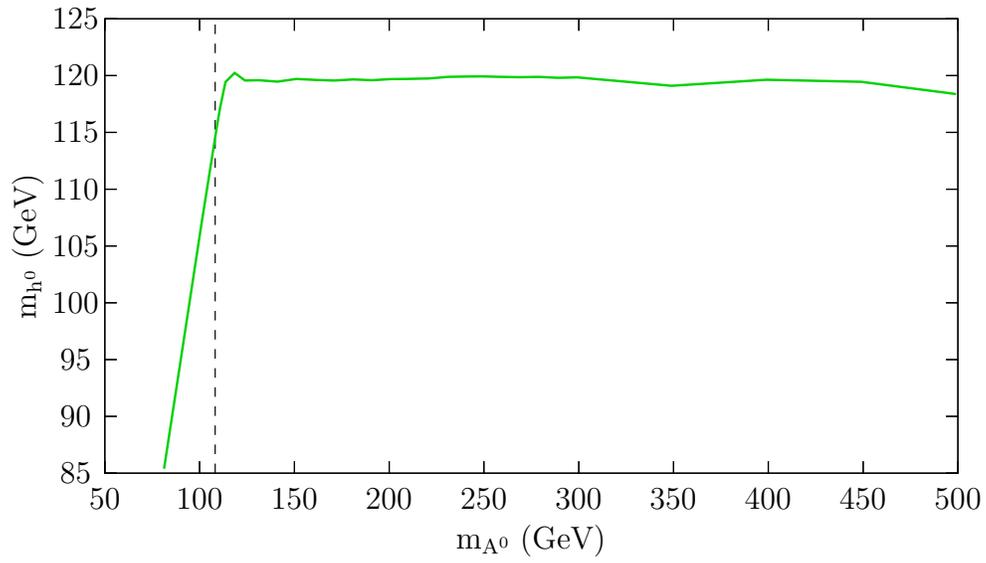}
\end{center}
\caption{Light Higgs mass  $m_h^0$ as a function of $m_A^0$.}
\end{figure}

\begin{figure}
\begin{center}
\input{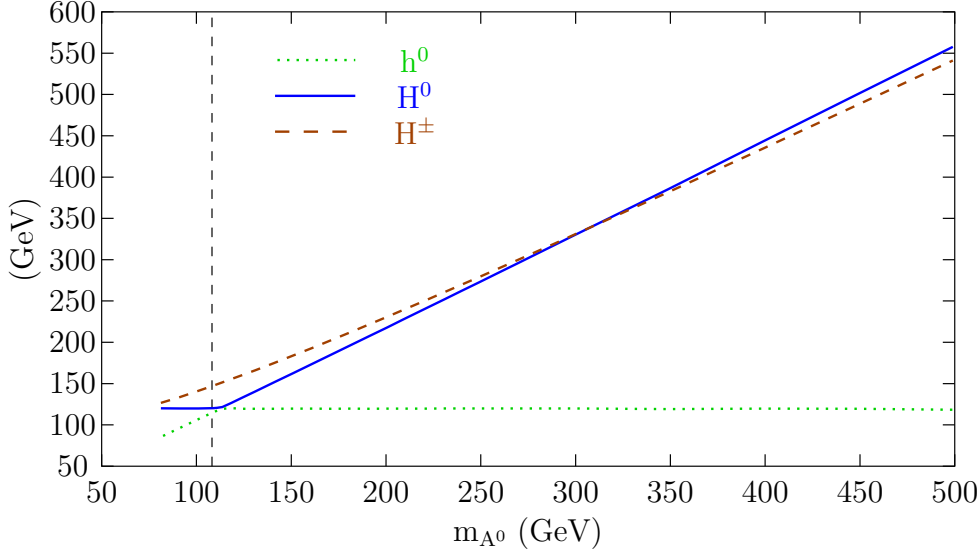}
\end{center}
\caption{Heavy $m_H^0$, light $m_h^0$ and charged Higgs $m_H^\pm$
 masses as a function of $m_A^0$.}
\end{figure}

\newpage

\section{Results: D term Higgs splitting}
\label{sect:dterm}
D term splitting for Higgs up/down masses seems natural.
We have thus performed a \ch2 analysis with D term splitting.   We find
that Yukawa unification does not work in this case, i.e. the best \ch2
obtained is $> 5$.   It is easy to understand why D term splitting does
not work.  $D_X > 0$, needed for electroweak symmetry breaking, makes
$m^2_{\bar d} < m^2_{\bar u}$ already at $M_G$ (see Eqn. \ref{eq:dterm}).
Therefore, in this case $m_{\tilde b_1} \leq m_{\tilde t_1}$ and hence the
gluino contribution to $\Delta m_b$ dominates.

It is important to check whether small threshold corrections to Yukawa unification can change this result.   In this analysis we take
 $|\epsilon_b|, |\epsilon_t|$ as large as 10\%\footnote{We do not consider threshold corrections to Yukawa unification in excess of 3\% to be small.
Such significant corrections would require additional physics explanations.}
in order to overlap with the
parameter range considered in a recent paper~\cite{baerferrandis}.
In Fig. 19  we plot constant \ch2 contours as a function of $\epsilon_b, \epsilon_t$ for $m_{16} = 2000$\gev, $\mu = 150 \; {\rm GeV}, \;
M_{1/2} = 300 \; {\rm GeV}$ fixed and all other input parameters
varied to minimize  \ch2.  Good fits are obtained with $\epsilon_t
\approx 0$ and $\epsilon_b \sim $ -7\% or with $\epsilon_t \sim -
\epsilon_b \sim $ 5\%.
These GUT scale corrections to Yukawa unification
are significant.   They are needed in this case for the RG evolution from
$M_G$ to $M_Z$ to drive $m_{\tilde t_1} << m_{\tilde b_1}$.  This is because
Yukawa couplings via the RGEs tend to drive squarks lighter, hence
$\lambda_t > \lambda_b$ compensates for the unfavorable boundary condition with  $m^2_{\bar d} < m^2_{\bar u}$.    Good fits are again obtained in the same narrow region of soft SUSY breaking parameters with $A_0 \sim
- 1.9 \; m_{16}$ and $ m_{10} \sim 1.35 \; m_{16}$, as in the case of Just So Higgs splitting and exact Yukawa unification (see Table 2).
A few comments are in order.  Neutrino threshold corrections give $\epsilon_b > 0$, assuming the sterile neutrinos obtain a Majorana mass $< M_G$ (see Appendix).  In addition, although $\epsilon_t \neq 0$ is $SU(5)$ invariant,
$\epsilon_b \neq 0$ requires  $SU(5)$ breaking threshold corrections.

\begin{figure}
\begin{center}
\input{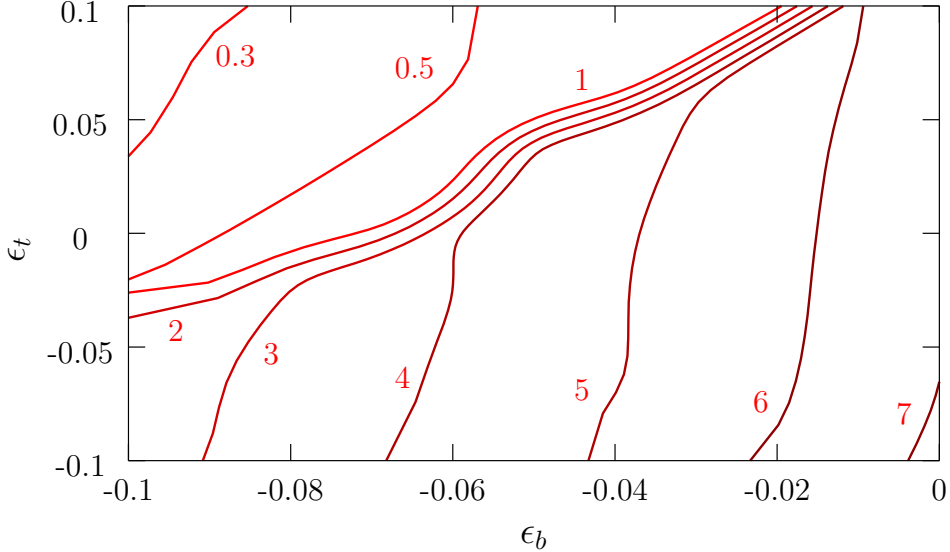}
\end{center}
\caption{\ch2 contours as a function of $\epsilon_b, \epsilon_t$
for $M_{1/2} = 300 \;{\rm GeV},\; \mu = 150 \; {\rm GeV}, m_{16} = 2 \;{\rm
TeV}$.   D term splitting case.}
\end{figure}

In Table 1 (Fit 4) we present our best fit point with D term splitting
and exact Yukawa unification.   This best fit has \ch2 $> 5$ and is
thus unacceptable.  Fit 5 on the other hand,  is one point in parameter
space with D term splitting and significant threshold corrections
to Yukawa unification, i.e. $\epsilon_b = -0.08, \;\; \epsilon_t = 0.02$.
The results are comparable to Fit 2 with Just So  Higgs splitting and
exact Yukawa unification.
This point with fixed  $m_{16} = 2000$\gev, $\mu = 150, \; M_{1/2}= 300$
is similar to but not completely consistent with the results of Ref.~\cite{baerferrandis}.
Besides the fact that we always have larger values for the GUT scale Yukawa
coupling than found in Ref.~\cite{baerferrandis}, we also require significantly smaller Yukawa threshold corrections.

\section{Experimental Tests \& More Constraints}
\label{sect:tests}

The most unexpected result of this analysis is the constraint on the light
Higgs mass.  We find $m_h^0 \sim  114 \pm 5 \pm 3$ GeV where the first
uncertainty comes from the range of SUSY parameters with \ch2 $\leq 1.5$
and the second is an estimate of the theoretical uncertainties in our
Higgs mass.  Surely this prediction will be tested at either Run III at the
Tevatron or at LHC.  We have used the analysis of Ref. \cite{carenaetal} which
is a good approximation for $m_t \; A_t/M_{(SUSY)}^2 < 0.5$ with $ M_{(SUSY)}^2 = (m_{\tilde t_1}^2 + m_{\tilde t_2})^2/2$.  In Fig. 20 we plot this ratio.  Note, due to our large value for $A_t$ we typically find $m_t \; A_t/M_{(SUSY)}^2 \sim 0.7$  which is somewhat outside the preferred range.
We have also compared our results with FeynHiggsFast and find ours
to be larger by about 3 GeV.   We are not certain of the reason for this
difference.

\begin{figure}
\begin{center}
\input{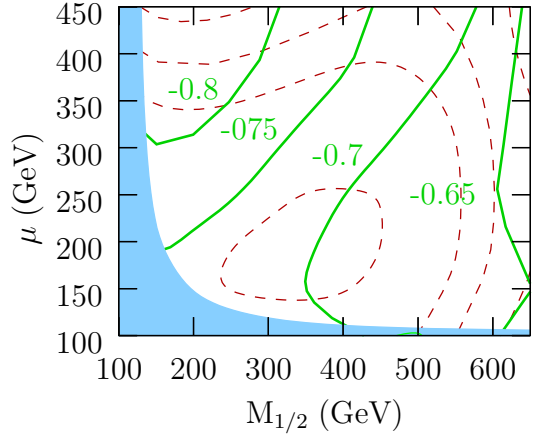}
\end{center}
\caption{Contours of constant $m_t A_t/M_{(SUSY)}^2$ with fixed $
m_{16} = 2000$ GeV, as a function of $\mu, \; M_{1/2}$.}
\end{figure}

Yukawa unification alone prefers light $A^0,\; H^0$ and $H^\pm$ masses.
However in the region of large $\tan\beta$ it has been shown that the process $B_s \rightarrow \mu^+ \; \mu^-$ provides a lower bound on $m_{A^0} \geq 200$ GeV (as pointed out in the recent works Dedes et al. and also Isidori and Retico~\cite{babukolda}) so that it is below the experimental upper bound $B(B_s \rightarrow \mu^+ \; \mu^-) < 2.6 \times 10^{-6}$ (95\% CL) \cite{abe}.~\footnote{We thank K.S. Babu and C. Kolda for discussions.}  This important new constraint is a consequence of the flavor violating quark-quark-$A^0$ couplings which result from the
large threshold corrections to CKM angles obtained in the region of large
$\tan\beta$ \cite{bpr}.  This has only a minor impact on $\chi^2$ as
discussed above.  We find that $\chi^2$ increases by at most 40\% for any $m_{A^0}$ less than $\approx 350$ GeV (Fig. 16).    The light Higgs mass $m_h^0$ is rather insensitive to the value of $m_{A^0}$  (Fig. 17); whereas $m_H^0, m_{H^+}$ are linearly dependent on $m_{A^0}$ (Fig. 18).    We are thus not able
to predict the other Higgs masses.  Direct observation of $A^0,\; H^0,\; H^\pm$
may be difficult at the Tevatron, but should be possible at LHC.  On the other
hand, $B_s \rightarrow \mu^+ \; \mu^-$ is a significant constraint and may be discovered at the Tevatron.

We find that the process $b \rightarrow s \gamma$ also provides significant
new constraints on SUSY parameter space.
In order to calculate $b \rightarrow s \gamma$ we have included second
family data ($m_s(2 \; {\rm GeV}) = 110 \pm 35 \; {\rm MeV},$ $ M_b - M_c = 3.4 \pm 0.2 \; {\rm GeV}, $ $ V_{cb} = 0.0402 \pm 0.0019$) in the \ch2 function
in order to self-consistently obtain flavor violating SUSY contributions.   We have used a parametrization of the Yukawa couplings at $M_G$ which, though not completely general, fits the data well (see Appendix).\footnote{Note, it has been shown (Ref. \cite{br}) that 2nd - 3rd family flavor mixing can significantly affect the result for $B(b \rightarrow s \gamma)$.}  We find that the coefficient $C_7^{MSSM}$ is of order $- C_7^{SM}$ (see for example, Eqn. 9 in Ref. \cite{br}) with the chargino term dominating
by a factor of about 5 over all other contributions.   This is due to the
light stop $\tilde t_1$.   In fact, $b \rightarrow s \gamma$ is more sensitive
to $m_{\tilde t_1}$ than $m_b(m_b)$.  This is because the amplitude
depends on the inverse fourth power of the stop mass while chargino
correction to the bottom mass depends only on the inverse second power.
 Fitting the central value
$B(b \rightarrow s \gamma) = 2.96 \times 10^{-4}$~\cite{gambinomisiak}
requires a heavier $\tilde t_1$ with
$(m_{\tilde t_1})_{MIN} \sim 450$ GeV; significantly larger than the range which
provides the best fits to $m_b$. Nevertheless, $\tilde t_1$ is still the lightest squark with significant stop-sbottom splitting. The $\tilde b_1, \; \tilde \tau_1$ masses also increase significantly.
We now find $m_b(m_b)_{MIN} \sim 4.3$.

In Table 2 we present representative points which are consistent with both $B(b \rightarrow s \gamma)$ and $B_s \rightarrow \mu^+ \; \mu^-$.  Fits 1 and 2 are with Just So  Higgs splitting, while Fits 3 and 4 are with D term splitting.  Fits 1,3,4 (Table 2) correspond to the same values of
$\mu,\; M_{1/2},\; m_{16}$ as Fits 2,4,5 (Table 1).  Fits 1 to 3 have exact Yukawa coupling unification at $M_G$.  Fit 4 has $\epsilon_b = -0.08, \;
\epsilon_t = 0.02$.   $B(b \rightarrow s \gamma)$ is the strongest
constraint in these fits.  Better agreement between Yukawa unification
and  $B(b \rightarrow s \gamma)$ is achieved with increasing $m_{16}$.
Also when we fit $B(b \rightarrow s \gamma)$ at $+3 \sigma$, then we obtain slightly lower squark and slepton masses with the changes indicated
in parentheses in Table 2.   

We have not reevaluated \ch2 contours including the $B_s \rightarrow \mu^+ \; \mu^-$ and $b \rightarrow s \gamma$ constraints.   We do not expect the \ch2
contours in the $\mu,\; M_{1/2}$ plane to change significantly, since we
can accomodate these new constraints with small changes in the parameter $A_0$
and negligible changes in all others.
Thus we expect that the predictions for gaugino masses to be unaffected by the $B_s \rightarrow \mu^+ \; \mu^-$ and $b \rightarrow s \gamma$ constraints.  Hence the lightest neutralino is the LSP
and a dark matter candidate~\cite{leszek}.   In order to know how observable
neutralinos and charginos may be,  we encourage the analysis of some new benchmark points consistent with Yukawa unification.

Finally, we recall that proton decay
experiments prefer values of $m_{16} > 2000$ GeV and $m_{16} >> M_{1/2}$ (see Ref. ~\cite{pdecay}).   This is in accord with the range of SUSY
parameters found consistent with third generation Yukawa unification.
There is however one experimental result which is not consistent with
either Yukawa unification or proton decay and that is the anomalous magnetic
moment of the muon.  Large values of $m_{16} \geq 1200$ GeV lead to very
small values for  $a^{NEW}_\mu \leq 16 \times 10^{-10}$.\footnote{Although $a^{SUSY}_\mu$ has not been included in the \ch2 function, we have included
the calculated values for $a^{SUSY}_\mu$ for the points in Tables 1 and 2.} Hence a necessary condition for Yukawa unification is that forthcoming BNL data~\cite{muon} and/or a reanalysis of the strong interaction contributions
to $a_\mu^{SM}$ will significantly decrease the discrepancy between the data
and the standard model value of $a_\mu$.

In summary, most of the results of our analysis
including only third generation fermions remain intact when incorporating
flavor mixing.  The light Higgs mass and most sparticle masses receive only
small corrections.  The lightest stop mass increases, due to
$b \rightarrow s \gamma$.  Nevertheless there is still a significant
$\tilde t_1 - \tilde t_2$ splitting and $m_{\tilde t_1} <<
m_{\tilde b_1}$.  The $A^0, \; H^0, \; H^+$ masses are necessarily
larger in order to be consistent with $B_s \rightarrow \mu^+ \; \mu^-$~\cite{babukolda}, which suggests that this process should be observed soon;
possibly at Run III of the Tevatron.
Finally, the central value for $a_\mu^{NEW}$ must significantly decrease.
The ``smoking guns" of SO(10) Yukawa unification, presented in this
paper, should be observable at Run III of the Tevatron or at LHC.  Also,
in less than a year we should have more information on $a_\mu^{NEW}$.

\section{Discussion}
\label{sect:discussion}

In the previous section we presented some experimental tests of Yukawa unification.  Here we consider some open theoretical questions.

Yukawa unification only works in a narrow region of SUSY parameter
space with $A_0 \approx - 1.9 \; m_{16}$, $m_{10} \approx 1.35 \; m_{16}$
and $m_{16} > 1200$ GeV.  The question arises, is this boundary condition natural in any SUSY breaking scheme?  In mSUGRA,  dilaton and
anomaly mediated SUSY breaking schemes $A_0 \neq 0$ at the GUT scale.
On the other hand, in gauge mediated or gaugino mediated SUSY breaking
schemes,  $A_0 = 0$ at $M_G$.   The latter are thus disfavored by
Yukawa unification.   In addition, anomaly mediated SUSY breaking has
other problems.  Slepton masses squared are negative unless other physics is added.  More importantly, however, since the gluino and chargino masses have opposite sign, it is difficult to simultaneously fit \bsgam and $a_\mu^{NEW}$.
Finally although $A_0 \neq 0$  at $M_G$ in mSUGRA and dilaton SUSY breaking
schemes,  this still does not explain why  $A_0 \approx - 1.9 \; m_{16}$.
However, the other relations for Higgs masses, i.e.  $m_{10} \approx 1.35 \; m_{16}$ and \dltmh = .13, may be obtained via RG running above $M_G$ or via threshold corrections.

It is an interesting, but not too surprising, result that the region of
SUSY parameter space preferred by Yukawa unification is very similar
to the region of SUSY parameter space needed to obtain heavy
1st and 2nd generation squarks and sleptons with third generation
squarks and sleptons lighter than O(TeV)~\cite{scrunching}.\footnote{Note,
this region of parameter space is desirable for suppressing
flavor and CP violating SUSY loops.}
First of all, the $SO(10)$ boundary conditions with $m^2_{(Q,\; \bar u,\; \bar d,\; L,\; \bar e)} = m_{16}^2$ and $m^2_{(H_u,\; H_d}) = m_{10}^2$ are obtained as a result of demanding the inverted scalar hierarchy in Ref. \cite{scrunching}, whereas for us they are input.
In addition, we need a light stop with large stop-sbottom splitting forcing us
to the same region of parameter space with large negative $A_0 \sim - 2 \; m_{16}$ and large $m_{16}$ with $m_{10} \sim \sqrt{2} \; m_{16}$.

It would be interesting to see how sensitive our results may
be to alternative electroweak symmetry breaking approximations.  In this
paper we have used the effective 2 Higgs doublet analysis of \cite{carenaetal}.
This approximation may be particularly well suited to the light Higgs
spectrum we obtain in our analysis.  The alternate scheme, in which
the Higgs tadpoles are evaluated in the MSSM at a scale of order $M_{stop}$
~\cite{pierceetal}, is however more frequently used in the literature.

We have a neutralino LSP and it is important to know if it is consistent
with cosmology and possibly a good dark matter candidate.  We note
that Yukawa unification places us in a region of SUSY parameter space which is
markedly different than has been studied in the literature.  A preliminary
investigation suggests that there are no major problems~\cite{leszek}.
Further study in this region of parameter space is now highly motivated
by our results.

Finally, let us consider the issue of vacuum stability.   Since we have
large $A_t$,  we may find that the vacuum stability condition
$|A_t| < m_{\tilde t_L} + m_{\tilde t_R}$~\cite{stability} in the
stop-Higgs sector is violated.   Indeed we find only a narrow region in
SUSY parameter space with \ch2 $< 1$ where this constraint is satisfied.
However the small change in $A_0$, necessary to fit $B(b \rightarrow s \gamma)$, is also sufficient to satisfy this stability constraint.

{\bf Acknowledgements}

S.R. and R.D. are partially supported by DOE grant DOE/ER/01545-826.
S.R. graciously acknowledges the Alexander von Humboldt Foundation, the Universit\"at Bonn and the Theory Division at CERN for partial support.
R.D. and S.R. thank the Physics Department at the Universit\"at Bonn.  We
also thank
the Theory Division at CERN for their kind hospitality while working on this
project.   Finally, R.D. and S.R. thank A. Dedes for his help and collaboration
in the early stages of this analysis.

\appendix
\section{Appendix}
In this Appendix we consider the possible GUT scale threshold corrections
to gauge and Yukawa unification and to soft SUSY breaking scalar masses.

\subsection{Gauge coupling unification}

At tree level the three gauge couplings unify at the GUT scale.
At one loop all three gauge couplings receive corrections depending logarithmically on an arbitrary scale $\mu$ and the masses of the
particles integrated out of the theory.   We may choose $\mu \equiv
M_G$ such that $\alpha_1(\mu) = \alpha_2(\mu) \equiv \alpha_G$.
Then the one loop threshold correction corresponds to a shift given by
$\epsilon_3 \equiv  (\alpha_3(M_G) - \alpha_G)/\alpha_G$.
$\epsilon_3$ obtains contributions from all massive states with SO(10)
quantum numbers.   We obtain~\cite{lucas}

\be {\epsilon}_3=f(\zeta_1,\ldots,\zeta_m) +
\frac{3 \tilde\alpha_G}{5\pi}\log \abs{{{\det}\bar M_t  \over M_G {\det} \bar M'_d}} +
\cdots \label{eq:ep3} \ee
where the first term represents the contributions from
$W_{sym\, breaking}$. It is only a function of U(1) and R invariant products of powers of vevs \{$\zeta_i$\}.  It is typically large O($\pm$10\%).  The second term comes from the Higgs sector where the color triplet and doublet mass
matrices $\bar M_t,\; \bar M'_d$ only include those states, from $5$s and $\bar 5$s of SU(5) contained in $W_{sym\, breaking}$ and $W_{Higgs}$, which mix with the Higgs sector.  For further details, see Ref. \cite{lucas}.

\subsection{Yukawa coupling unification}

The third generation Yukawa couplings are derived from the minimal
interaction  $W =  \L \; {\bf 16_3 \; 10_H \; 16_3}$.  At tree level
we have $\lambda_t = \lambda_b = \lambda_\tau = \lambda_{\nu_\tau}
\equiv \L$.   At one loop this relation is corrected~\cite{wright}.
However in this
case there are just three sources of corrections:  gauge exchange for
${\bf 10_H}$ and ${\bf 16_3}$; Yukawa exchange with color triplet Higgs
fields and with heavy right-handed neutrinos in the loop.
When one
considers a theory of fermion masses for three families then additional
Yukawa couplings mixing ${\bf 16_3}$ with other heavy SO(10) states
are possible.  However even if these new Yukawa couplings
are of order one, their contribution to threshold corrections are
typically less than 1\%.  More importantly however there are corrections which
come at tree level in the effective low energy field theory.   In a three
family model $3 \times 3$ Yukawa matrices are needed in order to obtain
both fermion masses and CKM mixing angles.    Upon diagonalizing the
Yukawa matrices, one effectively obtains tree level corrections to
the simple SO(10) relation.

\subsubsection{One loop corrections neglecting $\bar \nu_\tau$}

Consider first the one loop corrections neglecting $\bar \nu_\tau$.
We find
\be  \epsilon_b = \frac{\alpha_G}{2\pi} [\log (\frac{M_5^2}{M_G^2}) - 1]
- \frac{\L^2}{32 \pi^2} [\log (\frac{M_T^2}{M_G^2}) - 1]
\ee
\be  \epsilon_t = \frac{3 \alpha_G}{10 \pi}
[\log (\frac{M_{10}^2}{M_G^2}) - 1]
+ \frac{\alpha_G}{4 \pi} [\log (\frac{M_5^2}{M_G^2}) - 1]
\ee
where  $M_{10},\; M_5 \;\; (M_T)$ is the mass of the $SO(10)$ gauge fields contained in $SO(10)/SU(5)$, $SU(5)/(SU(3)\times SU(2)\times U(1))$ (the mass of the color triplet Higgs fields contained in $10_H$).
For $M_5 \approx M_T \sim M_G$ and $M_{10} \leq 10 M_G$ we obtain
$ \epsilon_b \sim - 0.7$\%, $ \epsilon_t \leq 1.1$\%.

\subsubsection{One loop corrections due to $\bar \nu_\tau$ alone}

The superpotential for the neutrino sector is given by
\be W = \L \; H_u \; L_\tau \; \bar \nu_\tau  +  \frac{1}{2} M_{\bar \nu_\tau}
\; \bar \nu_\tau \; \bar \nu_\tau  \label{eq:taubar}
\ee
where $M_{\bar \nu_\tau}$ is the effective Majorana mass for the right handed
tau neutrino.   Neutrino oscillations consistent with atmospheric neutrino data
suggest $M_{\bar \nu_\tau} \approx m_t^2/\sqrt{3.5 \times 10^{-3}\; {\rm eV^2}}
\approx 10^{13} \; {\rm GeV}$  or  $M_{\bar \nu_\tau}/M_G \approx 10^{-3}$.

Integrating out the right handed tau neutrino leads to equal finite wave function renormalization of $H_u, \; L_{\tau}$.  Hence
\be  \epsilon_b = \frac{\L^2}{32 \pi^2} [\log (\frac{M_G^2}{M_{\bar \nu_\tau }^2}) + 1], \;\;\; \epsilon_t = 0
\ee
or $\epsilon_b = 2.6$\% for $\L = 0.7, \; M_G = 3 \times 10^{16} \; {\rm GeV},
\; M_{\bar \nu_\tau} = 10^{13} \; {\rm GeV}$.    This is a considerable
correction which actually goes in the wrong direction for fitting
the third generation masses.   We discuss the sensitivity of our results
to such a correction in the text.

\subsubsection{Tree level corrections due to more realistic Yukawa matrices}

A significant threshold correction to Yukawa unification can come when
Yukawa matrices for three families are considered.   As an example,
in Table 2 the process $b \rightarrow s \; \gamma$ was calculated using the following ansatze for two family Yukawa matrices, since it has been
shown that they provide a good fit for fermion masses and mixing angles~\cite{brt}.

\begin{eqnarray}
Y_u =&  \left(\begin{array}{cc}
                  \epsilon^\prime & - r \; \epsilon     \\
                  r \; \epsilon & 1 \end{array} \right) \; \lambda &
\nonumber \\
Y_d =&  \left(\begin{array}{cc}
                  \epsilon &  -  \sigma \;r \; \epsilon\\
                r\;  \epsilon & 1 \end{array} \right) \; \lambda
& \label{eq:yukawa}
  \\
Y_e =&  \left(\begin{array}{cc}
                   3 \; \epsilon  &  3 \; r\; \epsilon \\
                   - 3\; \sigma \;r \; \epsilon & 1 \end{array} \right) \; \lambda &
 \nonumber
\end{eqnarray}
The universal Yukawa coupling  $\lambda$ plus the three new complex parameters $\epsilon, \epsilon^\prime \;{\rm and}\; \sigma$ and one real parameter $r$
were varied to minimize the \ch2 function with five additional observables
$M_\mu = 105.66 \pm 1.1$ MeV, $M_b - M_c = 3.4 \pm 0.2$ GeV, $m_s(2 \; {\rm GeV}) = 110 \pm 35$ MeV, $V_{cb} = 0.0402 \pm 0.0019$ and
$B(b \rightarrow s \gamma) \times 10^3 = 0.296 \pm 0.035$.
Clearly there are more parameters than
observables, but we are not attempting in this analysis to make any new
predictions for fermion masses.  We just want to be able to calculate
the branching ratio for $b \rightarrow s \; \gamma$ self-consistently.
With Fit 1 in Table 2 as a guide, we find (upon diagonalizing the $2 \times 2$ Yukawa matrices at the GUT scale) $\epsilon_b \approx \epsilon_t \sim -0.08$.
Several points should be made here.  The first is that this tree level correction is model dependent and much larger than any one loop correction.
Secondly, $\epsilon_b \neq 0$ is a consequence of the Georgi-Jarlskog like
mass matrices distinguishing quarks and leptons.  It is needed to obtain a reasonable fit to $m_s(2 \; {\rm GeV})$ and $M_\mu$.  Finally, since $V_{cb} \sim 0.04$ is small, we find $\epsilon_t \approx \epsilon_b$.  This latter result is in the wrong direction for obtaining good fits to third generation masses with D term splitting as is evident in Fig. 19 and Fit 4 (Table 2).

\subsection{Higgs mass splitting}

D term splitting of the $H_u, \; H_d$ masses is quite natural in
$SO(10)$ SUSY GUTs.   It can be generated in the process of
breaking $SO(10) \rightarrow SU(5)$ by a mismatch
in the vacuum expectation values of the {\bf 16} and ${\bf \overline{16}}$
which are needed to break $SO(10) \rightarrow SU(5)$ and reduce the
rank of the group.   Once this D term is generated it then gives mass
to scalars proportional to their $U(1)_X$ charge.

The Just So case does not at first sight appear to be
similarly well motivated.   In this Appendix we attempt to rectify this
apparent difficulty.   It is quite clear that in any SUSY model the
Higgs bosons are very special.  R parity is used to distinguish
Higgs from squarks and sleptons.   In addition, a supersymmetric mass term
$\mu$ with value of order the weak scale is needed for the Higgs bosons.
Since $\mu$ is naturally of order $M_G$, one needs some symmetry argument
why it is suppressed.  Of course, if the Higgs are special, then perhaps this
will help us understand how to obtain splitting of the Higgs up/down
while maintaining universal squark and slepton masses.   GUT threshold corrections to soft SUSY breaking scalar masses have been considered
previously.  In Murayama et al.~\cite{nothreshcorr} it was shown that the necessary condition, $m_{10} > m_{16}$, can naturally be obtained in $SO(10)$
with RG running from $M_{Pl}$ to $M_G$.  In the paper by Polonsky and Pomarol~\cite{ewsb} the splitting of the soft masses of $SU(5)$ multiplets
within irreducible $SO(10)$ representations was considered.  In the following
we consider two novel sources for Higgs up/down splitting in the
context of $SO(10)$.

\subsubsection{$\nu_\tau$ contribution to Higgs splitting}

In the MSSM superpotential below $M_G$ we have the $\nu_\tau$ contribution
which distinguishes $H_u$ and $H_d$ (see Eqn. \ref{eq:taubar}).
This leads to a significant threshold correction
\be \Delta m_{H_u}^2 \approx  \frac{\L^2}{16 \pi^2} \; (2 m_{16}^2 + m_{10}^2
+ A_0^2) \log (\frac{M_{\bar \nu_\tau}^2}{M_G^2})
  + \; {\rm non \; log \;\; terms}
\ee
Using the values $\L = 0.7,\; M_{\bar \nu_\tau} = 10^{13} \; {\rm GeV}$ and $M_G = 3 \times 10^{16} \; {\rm GeV}$ and the typical boundary conditions
$A_0^2 \approx 2 \, m_{10}^2 \approx 4 \, m_{16}^2$, 
we obtain \dltmh $\equiv \frac{1}{2} \, \Delta m_{H_u}^2/m_{10}^2 = .10$.   Note this is remarkably close to the value needed for Just So Higgs splitting (see Fig. 5 (Right) and Fits 1 - 3 (Table 1) and Fits 1,2 (Table 2)).

\subsubsection{Another possible source for Higgs splitting}

Consider also the possible superpotential for the Higgs sector
\be  W =  \L_A \; 10 \; 45 \; 10^\prime \; + \; X \; (10^\prime)^2 \;
 + \; \psi \; \psi \; 10^\prime \; + \; M {\bar \psi} \psi \; + \; Tr(45^2 {\cal M}_{45}^2)
\ee
The first two terms are necessary for Higgs doublet/triplet splitting with
assumed vevs for the adjoint field $\la 45 \ra \sim (B - L)\; M_G$ and
$\la X \ra << M_G$.   The fields $\psi, \; {\bar \psi}$ are in $ 16, \overline{16}$ dimensional representations.  Their vevs break $SO(10)$
to $SU(5)$ and split the masses of $5_{10^\prime},\;\;
\bar 5_{10^\prime}$.  With this splitting and also an assumed $SU(5)$
invariant splitting
in the supersymmetric masses for the $45$ we find significant threshold corrections to \dltmh.

Schematically, we find
\be \Delta m_H^2 \propto   \frac{\L_A^2}{16 \pi^2} \log({\cal M}_{10_{45}}^2/{\cal M}_{24_{45}}^2)
 \; sin(\theta + \theta^\prime) sin(\theta - \theta^\prime) \{ r \}
 + \cdots   \ee
The factor $r$ represents the ratio of soft scalar masses
$m_{45}^2/ m_{10}^2,\;\;
m_{10^\prime}^2/ m_{10}^2,\;\; A_0^2/ m_{10}^2$
and finally the dots represent the contribution
of color triplet states to the loops.
The correct sign for \dltmh can always be obtained.
If $\L_A^2/4 \pi \sim O(1)$ or $r >> 1$, then we can easily obtain \dltmh $\sim 20 - 30$\%.  In the latter case, top, bottom unification will have
considerably smaller threshold corrections.

%

\newpage

\protect
\begin{table}
\caption[8]{
{\small Five representative points of the fits.  The first three are with
Just So Higgs splitting and the last two are with D term splitting.
All fits assume exact Yukawa unification at the GUT scale, except for
5 which has $\epsilon_b = -0.08, \;\; \epsilon_t = 0.02$.
All fits are with one loop RG running from $M_G$ to
$M_Z$ for dimensionful parameters, except for 3 which uses two loop running.
All entries are in units of GeV to the appropriate power.}
}
\label{t:table1}
$$
\begin{array}{|l|c|c|c|c|c|c|}
\hline
{\rm Data \; points}  & &  1 &  2  & 3 & 4 & 5 \\
\hline
 {\rm Input\; parameters}  & &  & & & & \\
\hline
\;\;\;\alpha_{G}^{-1} & & 24.46 &  24.66  & 24.66 & 24.73 & 24.58  \\
\;\;\; M_G \times 10^{-16} & & 3.36  &  3.07  & 3.07 & 3.13 & 3.16 \\
\;\;\;\epsilon_3 & & -0.042&  -0.040 & -0.040 & -0.046 &
-0.039 \\
\;\;\;\L  & &  0.70  & 0.67 & 0.67 & 0.80 & 0.63 \\
\hline
\;\;\; m_{16} & & 1500 &  2000 & 2000 & 2000 & 2000 \\
\;\;\; m_{10}/m_{16} & & 1.35 &  1.35 & 1.35 & 1.20 & 1.33 \\
\;\;\; \Delta m_H^2  & &   0.13 &  0.13 & 0.12 &  0.07 & 0.05 \\
\;\;\; M_{1/2} & & 250 &  350 & 350 & 350 & 300 \\
\;\;\;\mu & & 150 & 200 & 200 & 115 & 150 \\
\;\;\; \tan\beta & & 51.2  &  50.5 & 50.6 & 54.3 & 51.1 \\
\;\;\; A_0/m_{16} &  & -1.83 &  -1.87 & -1.83 & -0.37 & -1.87 \\
\hline
\hline
\chi^2 \; {\rm observables} & {\rm Exp}\;(\sigma)  &    &  & & & \\
\hline
\;\;\;M_Z & 91.188 \;(0.091) & 91.13 &  91.14 & 91.14 & 91.15 & 91.15 \\
\;\;\;M_W & 80.419 \;(0.080) & 80.45&  80.45 & 80.44 & 80.44 & 80.44 \\
\;\;\;G_{\mu}\times 10^5 & 1.1664 \;(0.0012) & 1.166 &  1.166 & 1.166 & 
1.166 & 1.166 \\
\;\;\;\alpha_{EM}^{-1} & 137.04\; (0.14) & 137.0 &  137.0 & 137.0 & 137.0 & 
137.0 \\
\;\;\;\alpha_s(M_Z)  & 0.118 \; (0.002)& 0.1175 &  0.1176 & 0.1175 & 0.1161 
& 0.1179 \\
\;\;\;\rho_{new}\times 10^3 & -0.200 \; (1.10) & 0.696& 0.460 & 0.437 & 
0.035 & 0.265 \\
\hline
\;\;\;M_t   &174.3\; (5.1)    & 175.5 &  174.6 & 174.4 & 177.9 & 174.1 \\
\;\;\;m_b(m_b) & 4.20 \; (0.20)  &  4.28  &    4.27 & 4.28 & 4.59 & 4.22 \\
\;\;\;M_{\tau} & 1.7770 \; ( 0.0018)  & 1.777 &  1.777 & 1.777 & 1.777 & 
1.777 \\
\hline
 {\rm TOTAL}\;\;\;\; \chi^2 &  & 1.50 & 0.87 & 0.91 & 5.42 & 0.45 \\
\hline
\hline
\;\;\; h^0      &        &  116  & 116 & 117 & 115 & 116 \\
\;\;\; H^0       &        &  120  & 121 & 121 & 117 & 120 \\
\;\;\; A^0        &       &  110  & 110 & 111 & 110 & 110 \\
\;\;\; H^+         &      &  148  & 148 & 149 & 146 & 148 \\
\;\;\; \tilde \chi^0_1 &  &  86 & 130 & 130 & 86 & 99 \\
\;\;\; \tilde \chi^0_2  & &  135  & 190 & 189 & 126 & 152 \\
\;\;\; \tilde \chi^+_1  & &   123 & 178 & 177 & 105 & 131 \\
\;\;\; \tilde g         &  &   661 &  913 & 898 & 902 & 787 \\
\;\;\; \tilde t_1       &  &  135  & 222 & 235 & 1020 & 141 \\
\;\;\; \tilde b_1       & &  433 & 588 & 589 & 879 & 342 \\
\;\;\; \tilde \tau_1    &  & 288   & 420 & 542 & 1173 & 693 \\
\hline
\;\;\; a_\mu^{SUSY} \times 10^{10}  & 25.6 \; (16)    & 9.7 & 5.5 & 5.5 & 
6.1 & 6.4 \\
\hline
\end{array}
$$
\end{table}

\newpage

\protect
\begin{table}
\caption[8]{
{\small  Four representative points of the fits.  The first two are with
Just So Higgs splitting and the last two are with D term splitting.
All fits assume exact Yukawa unification at the GUT scale, except for
4 which has $\epsilon_b = -0.08, \;\; \epsilon_t = 0.02$.
All fits are with one loop RG running from $M_G$ to
$M_Z$ for dimensionful parameters.
For these fits the branching ratio for $b \rightarrow s \gamma$ is included
in \ch2. In addition,  the CP odd Higgs mass $m_A^0$  is constrained to be 
200
GeV.   These points are thus consistent
with both $B(b \rightarrow s \gamma)$ and $B_s \rightarrow \mu^+ \; \mu^-$.
In Fits 1, 2 we show the change in the $\tilde t_1, \; \tilde b_1,\;
\tilde \tau_1$ masses if we fit $b \rightarrow s \; \gamma$ at the central
value + $3 \sigma$.}
}
\label{t:table2}
$$
\begin{array}{|l|c|c|c|c|c|}
\hline
{\rm Data \; points}  & & 1 & 2 & 3 & 4 \\
\hline
 {\rm Input\; parameters}  & &   &   & &  \\
\hline
\;\;\;\alpha_{G}^{-1} & & 24.72& 24.78 & 24.75 & 24.62  \\
\;\;\; M_G \times 10^{-16} & & 3.00 & 3.12 & 3.09 & 3.20 \\
\;\;\;\epsilon_3 & &  -0.040 & -0.041 & -0.045 & -0.043 \\
\;\;\;\L  & & 0.63 & 0.61 & 0.80 & 0.60 \\
\hline
\;\;\; m_{16} & &  2000 & 3200 & 2000 & 2000 \\
\;\;\; m_{10}/m_{16} &  & 1.32  &  1.30 & 1.19 & 1.30 \\
\;\;\; \Delta m_H^2  & & 0.13  &  0.13 & 0.07 & 0.05 \\
\;\;\; M_{1/2} &  &  350 & 350 & 350 & 300 \\
\;\;\;\mu & & 200 & 150 & 115  & 150 \\
\;\;\; \tan\beta & & 52.5 & 50.6 & 55.0 & 50.7 \\
\;\;\; A_0/m_{16} &  & -1.71 & -1.83 & -0.05 & -1.75 \\
\hline
\hline
\chi^2 \; {\rm observables} & {\rm Exp}(\sigma)  &    &   & &  \\
\hline
\;\;\;M_Z & 91.188 \; (0.091) & 91.18  & 91.18 & 91.16 & 91.18  \\
\;\;\;M_W & 80.419 \; (0.080) &  80.42   & 80.42 & 80.43 & 80.42 \\
\;\;\;G_{\mu}\times 10^5 & 1.1664 \; (0.0012) & 1.166 & 1.166 & 1.166 & 
1.166 \\
\;\;\;\alpha_{EM}^{-1} & 137.04\; (0.14) & 137.0  & 137.0 & 137.0 & 137.0 \\
\;\;\;\alpha_s(M_Z)  & 0.118 \; (0.002) & 0.1172  & 0.1173 & 0.1162 & 0.1167 
\\
\;\;\;\rho_{new}\times 10^3 & -0.200 \; (1.10) & 0.228 & 0.321 & 0.221 & 
0.279 \\
\hline
\;\;\;M_t   &174.3 \; (5.1)  & 173.8 &  172.1 & 178.8  & 172.6 \\
\;\;\;m_b(m_b) & 4.20 \; (0.20)  &  4.46 & 4.42 & 4.56 & 4.50 \\
\;\;\;M_{\tau} & 1.7770\; ( 0.0018)  &  1.777 &  1.777 & 1.777 & 1.777 \\
\hline
 {\rm TOTAL}\;\;\;\; \chi^2 &   & 2.05 & 1.72 & 5.03 & 3.38 \\
\hline
\hline
\;\;\; h^0      &        & 118  & 119 & 115 & 118 \\
\;\;\; H^0       &        & 217 & 217 & 218 & 216 \\
\;\;\; A^0        &      & 200 & 200 & 200 & 200 \\
\;\;\; H^+         &      & 229 & 229 & 228 & 228 \\
\;\;\; \tilde \chi^0_1 &  & 130 & 110 & 86 & 99 \\
\;\;\; \tilde \chi^0_2  & & 190 & 160 & 126 & 152 \\
\;\;\; \tilde \chi^+_1  & & 178 & 136 & 105 & 131 \\
\;\;\; \tilde g         &  & 909 & 904 & 902 & 781\\
\;\;\; \tilde t_1       &  & 509 \; (-30)& 511 \; (-27) & 1067 & 443 \\
\;\;\; \tilde b_1       & & 749 \; (-42) & 903 \; (-14) & 900 & 550 \\
\;\;\; \tilde \tau_1    &  & 459 \; (+47) & 1001 \; (-25) & 1173 & 854 \\
\hline
\;\;\; B(b \rightarrow s \gamma) \times 10^3 & 0.296 \; (0.035) & 0.297 &  
0.297 & 0.297 & 0.297 \\
\hline
\;\;\; a_\mu^{SUSY} \times 10^{10}  & 25.6 \; (16)  & 5.8 & 2.2 & 6.4 & 6.4 
\\
\hline
\end{array}
$$
\end{table}

\pagestyle{empty}

\end{document}